\documentclass[12pt]{article}
\pdfoutput=1
\usepackage{setspace}
\usepackage{amsmath}
\usepackage{amssymb}
\usepackage{colortbl}
\usepackage{gensymb}
\usepackage[table,x11names]{xcolor}
\usepackage{graphicx}
\usepackage{hyperref}
\usepackage{cite}
\usepackage{slashed}
\usepackage{pifont}
\usepackage{longtable}
\usepackage{multirow}
\usepackage{ltablex}
\usepackage{array}
\usepackage{rotating}
\usepackage{booktabs}
\usepackage{hhline}
\definecolor{nicered}{rgb}{0.6,0,0}
\definecolor{nicegreen}{rgb}{0.1,0.5,0.1}
\definecolor{niceblue}{rgb}{0,0.4,0.8}
\hypersetup{colorlinks,citecolor= nicegreen,linkcolor=black,urlcolor=niceblue}

\allowdisplaybreaks

\begin{document}
\begin{titlepage}
% -------------------------------------------------
% Uncomment to add preprint numbers when necessary
% -------------------------------------------------
% \begin{flushright}

% \end{flushright}

  \newcommand{\AddrLiege}{{\sl \small IFPA, D\'ep. AGO, Universit\'e de
      Li\`ege, B\^at B5, Sart Tilman B-4000 Li\`ege 1,
      Belgium}}
  \newcommand{\AddrUFSM}{{\sl \small  Universidad T\'ecnica 
      Federico Santa Mar\'{i}a - Departamento de F\'{i}sica\\
      Casilla 110-V, Avda. Espa\~na 1680, Valpara\'{i}so, Chile}}
  \newcommand{\AddrULB}{{\sl \small Universit\'e Libre de Bruxelles-Service 
      de Physique Th\'eorique CP225, Blvd. du Triomphe\\
      (Campus de la Plaine), B-1050 Bruxelles, Belgium}}
  \vspace*{0.5cm}
\begin{center}
  \textbf{\Large Neutrino non-standard interactions 
    and dark matter
    \\[3.5mm]
    searches with multi-ton scale detectors}
  \\[9mm]
  D. Aristizabal Sierra$^{a,b,}$\footnote{e-mail address: {\tt
      daristizabal@ulg.ac.be}},
  N. Rojas$^{a,}$\footnote{e-mail address: {\tt nicolas.rojasro@usm.cl}},
  M.H.G. Tytgat$^{c,}$\footnote{email address: {\tt mtytgat@ulb.ac.be}}
  \vspace{0.8cm}\\
  $^a$\AddrUFSM\\[3mm]
  $^b$\AddrLiege\\[3mm]
  $^c$\AddrULB
\end{center}
\vspace*{0.2cm}
\begin{abstract}
  \onehalfspacing
  Future dark matter (DM) direct detection searches will be subject to
  irreducible neutrino backgrounds that will challenge the
  identification of an actual WIMP signal in experiments without
  directionality sensitivity. We study the impact of neutrino-quark
  non-standard interactions (NSI) on this background, assuming the
  constraints from neutrino oscillations and the recent COHERENT
  experiment data, which are relevant for NSI mediated by light
  mediators, $m_{\rm med} \lesssim{\cal O}$(GeV). We calculate the
  expected number of neutrino-nucleus elastic scattering events in a
  Xe-based ton-size dark matter detector, including solar neutrino
  fluxes from the $pp$ chain and CNO cycle as well as sub-GeV
  atmospheric fluxes and taking into account NSI effects in both
  propagation and detection. We find that sizable deviations from the
  standard model expectation are possible, but are more pronounced for
  flavor-diagonal couplings, in particular for electron neutrinos. We
  show that neutrino NSI can enhance or deplete the neutrino-nucleus
  event rate, which may impact DM searches in multi-ton detectors.
\end{abstract}
\end{titlepage}
\setcounter{footnote}{0}
\tableofcontents
\vspace{1cm}
\section{Introduction}
\label{sec:intro}
A wealth of astrophysical and cosmological data support the idea that
the dominant form of matter in the universe is made of dark matter
(DM).  Several DM candidates have been proposed ranging from
astrophysical observable objects \cite{Griest:1990vu} up to
fundamental particles \cite{Feng:2010gw}. Among the latter, the most
studied candidates (both theoretically and experimentally) are massive
particles whose abundance is determined by thermal freeze-out (weakly
interacting massive particles or WIMPs).

WIMPs are being searched for using direct detection experiments since
the late 80's, following a proposal by Goodman and Witten
\cite{Goodman:1984dc}. These experiments rely on energy deposition
measurements from WIMP-induced nuclear recoil processes, which are
expected to result as the Earth (detector) moves in the galactic DM
halo. The non-observation of such signatures are translated into upper
limits for the WIMP-nucleon (spin-independent or spin-dependent) cross
section as a function of the DM mass. At present the most stringent
bound is given by the XENON1T collaboration, which in the absence of a
nuclear recoil signal has derived the limit
$\sigma_{\chi-n}\lesssim 7.7\times 10^{-47}\,\text{cm}^2/c^2$ for
$m_\chi=35\,\text{GeV}/c^2$ at 90\% CL \cite{Aprile:2017iyp}. Searches
beyond these values and down to, say,
$\sigma_{\chi-n}\sim 10^{-49}\,\text{cm}^2/c^2$ require multi-ton
target masses, as planned for the DARWIN detector
\cite{Aalbers:2016jon}.

The reach of these experiments, however, will be limited by an
irreducible neutrino background. Indeed, coherent neutrino nucleus
elastic scattering (C$\nu$NES) induced by solar and, eventually,
atmospheric neutrino fluxes will mimic a possible WIMP-nucleus signal
\cite{Billard:2013qya}. While the precise extent of this background is
affected by various factors, including neutrino flux uncertainties and
astrophysical parameters \cite{Billard:2013qya,OHare:2016pjy}, in the
SM, the C$\nu$NES cross section is well understood and is not a source
of uncertainty. This can potentially change in the presence of new
physics. The most obvious way through which this can occur is via
neutrino-quark non-standard interactions (NSI), which historically
were introduced by Wolfenstein as a possible way to account for the
solar neutrino deficit \cite{Wolfenstein:1977ue}.

Neutrino NSI are constrained by several observations, including
neutrino oscillation data \cite{Gonzalez-Garcia:2013usa}, neutrino
deep inelastic scattering experimental data (CHARM
\cite{Dorenbosch:1986tb} and NuTeV \cite{Zeller:2001hh}) and,
potentially, the measurement of the C$\nu$NES process itself, recently
observed by the COHERENT experiment \cite{Akimov:2017ade} (see
\cite{Farzan:2017xzy} for a review and \cite{Coloma:2017ncl} for the
detailed analysis). CHARM and NuTeV set strong constraints on
$\nu_\mu$-related NSI parameters, but their relevance for constraining
C$\nu$NES depends on the range of application of the NSI effective
operators. In particular, these constraints do not apply if the NSI
are mediated by a particle with mass $m_{\rm med} < {\cal O}(1)$ GeV
\cite{Coloma:2017egw}.  In what follows we will consider the
possibility that NSI are mediated by a particle in the mass range
$10\,\text{MeV} \lesssim m_{\rm med} \lesssim 1\,\text{GeV}$, range in
which CHARM and NuTeV bounds do not apply, but one is still allowed to
use the constraints on NSI as extracted from a combined analysis of
COHERENT and oscillation data in ref.\cite{Coloma:2017ncl}. It is
worth pointing out that in this mass range bounds are less severe for
all flavor-diagonal NSI couplings, and so compared to the case where
deep inelastic scattering data apply one expects larger deviations on
the neutrino-nucleus event rates.

Specifically, using current constraints on neutrino NSI derived from
neutrino oscillation and COHERENT data \cite{Coloma:2017ncl}, in this
paper we quantify the deviations on the C$\nu$NES expected number of
events in a Xe-based DM detector using an exposure
$\mathcal{E}=1\;(\text{ton}\cdot\text{year})$. For that aim we
consider solar neutrino fluxes for which we include neutrinos from the
$pp$ chain ($pp, hep, ^8\text{B}, ^7\text{Be}, pep$) and CNO cycle
($^{13}\text{N}, ^{15}\text{O}, ^{17}\text{F}$). We consider as well
sub-GeV atmospheric neutrino fluxes. For the solar sector analysis we
calculate NSI propagation effects in the mass dominance limit
($\Delta m_{31}^2\to\infty$) and for the calculation of the electron
neutrino survival probability we take into account the distribution of
neutrino production in the Sun according to the standard solar model
(SSM) BS05 \cite{Bahcall:2004pz}.  We quantify matter-induced effects
in the atmospheric sector and find they are negligible
($H_\text{mat}/H_\text{vac}\lesssim 0.1$) \cite{Friedland:2004ah},
however we include neutrino flavor oscillations in vacuum in the
calculation of the number of neutrino-nucleus events.

The study of beyond-the-standard model physics in neutrino-quark
interactions in the context of multi-ton DM detectors, has been
recently the subject of various analyses. Ref. \cite{Cerdeno:2016sfi}
studied the capability of such detectors to constraint light mediators
scenarios, while ref. \cite{Bertuzzo:2017tuf} instead considered
scenarios where the new physics couples both neutrinos and
DM. Ref. \cite{Dutta:2017nht} considered the implications of neutrino
NSI on neutrino backgrounds. The analysis was done for solar neutrinos
and in particular for the $^8$B neutrino flux, for which solar
neutrino fluxes can mimic a WIMP-nucleus scattering event. We extend
upon this study by reanalyzing NSI effects in the case of lighter
mediators, and so in the light of the recently released COHERENT
data. We furthermore consider NSI effects all over solar neutrino
energies and include sub-GeV atmospheric neutrino fluxes. The former
is particularly relevant if one aims at setting e.g. background-free
sensitivities for light WIMPS.

The rest of this paper is organized as follows. In sec. \ref{sec:nsi}
we introduce neutrino NSI, first assuming they are generated by heavy
mediators ($q^2\ll m_X^2$) \footnote{They might as well proceed
  through the exchange of light mediators ($q^2\gg m_X^2$). The
  results for this case will be presented elsewhere
  \cite{daristi:2017-c}.} and review current experimental allowed
ranges. In sec. \ref{sec:dm-detectors} we discuss C$\nu$NES in the
presence of NSI, calculate electron neutrino survival probabilities in
the Sun for various NSI data sets and study atmospheric vacuum
conversion probabilities. In sec. \ref{sec:nsi-dm-detectors} we
present our main results, namely the maximum and minimum number of
C$\nu$NES events in ton-size DM detectors (Xe based). We perform a
single-parameter analysis, \textit{c'est-\`a-dire} we assume all NSI
parameters are zero but one and vary that coupling within its
experimental allowed range. Although constrained, this analysis enable
us to capture the main features of what should be expected from
neutrino NSI in near-future DM direct detection experiments. Finally,
in sec. \ref{sec:concl} we summarize and present our conclusions.
\section{Neutrino-quark non-standard interactions}
\label{sec:nsi}
NSI can be parametrized in terms of the following effective Lagrangian
(see e.g. \cite{Ohlsson:2012kf}):
\begin{equation}
  \label{eq:eff-Lag-NSI}
  \mathcal{L}_\text{Eff}=-\sqrt{2}G_F\sum_{q=u,d}\overline{\boldsymbol{\nu}}
  \gamma_\mu P_L\left(\boldsymbol{\epsilon^{qV}} 
    + \boldsymbol{\epsilon^{qA}}\gamma_5 \right)\boldsymbol{\nu}
  \;\;\overline{q}\gamma^\mu q\ ,
\end{equation}
where $\boldsymbol{\nu}^T=(\nu_e, \nu_\mu, \nu_\tau)$ and
$\boldsymbol{\epsilon^{qV}}$ and $\boldsymbol{\epsilon^{qA}}$ refer to
$3\times3$ matrices in flavor space whose elements correspond to
vector and vector-axial couplings. Hereafter we will denote
flavor-related matrices in bold-face. This parametrization assumes
that the neutrino NSI are related to new physics at a scale
$m_\text{med}\gg q$, with $q$ the characteristic energy momentum
transferred in the relevant process.  It can be as well that the new
physics is related with a scale way below the typical momentum
exchange (light mediator scenario), in which case a phenomenological
analysis requires to take into account the dynamics of the
mediator. Such models with light mediators have been discussed in, for
instance \cite{Farzan:2015doa,Farzan:2015hkd,Babu:2017olk}. Here, we
focus on the intermediate possibility where CHARM and NuTeV
constraints do not apply, while an effective approach is still valid
so that the constraints on NSI parameters from the COHERENT data
analyzed in \cite{Coloma:2017ncl} hold. In this regime, the
constraints on the flavor-diagonal NSI couplings are substantially
weaker. For instance (see e.g. \cite{Farzan:2017xzy})
  \begin{equation}
    \label{eq:constraints-farzan}
    \vert \epsilon_{\mu\mu}^{dV}\vert < 0.042 \quad (\mbox{Atm. + Acc.})
    \longrightarrow -0.075 < \epsilon_{\mu\mu}^{dV}  < 0.33
    \quad (\mbox{Oscillation + COHERENT}) 
\end{equation}
Accelerator constraints are irrelevant for mediator masses below
${\cal O}(1$ GeV). The precise value is not important here, as long as
it is larger than about $100$ MeV, the maximal momentum exchange scale
we will consider.  Since \cite{Coloma:2017ncl} assumes
$m_{\rm med} \gtrsim 10 \mbox{\rm MeV}$, our analysis will be
therefore limited to
$10 \mbox{\rm MeV} < m_{\rm med} < 1 \mbox{\rm GeV}$. While \textit{a
  priori} this seems a narrow window in parameter space, it is however
for that window where the largest possible deviations---compared to SM
expectations---on the neutrino-nucleus scattering rate are expected.

The interactions in (\ref{eq:eff-Lag-NSI}) lead to forward coherent
scattering (order $G_F$ interaction) and scattering (order $G_F^2$
interaction) processes such as C$\nu$NES. The former are responsible
for matter potentials in the Sun and in the Earth and are entirely
controlled by $\boldsymbol{\epsilon^{qV}}$. For the latter, the vector
current determines spin-independent processes while the axial-vector
current accounts for spin-dependent ones.  Reactions rates driven by
vector-axial currents are relatively smaller (in particular for large
nuclei) and so we will not consider them. From now on then we drop the
index $V$.

Constraints on NSI are derived from neutrino oscillation
\cite{Bolanos:2008km,Escrihuela:2009up,Gonzalez-Garcia:2013usa,Farzan:2017xzy}
and COHERENT data \cite{Akimov:2017ade}. Let us discuss this in more
detail. The presence of the vector current in (\ref{eq:eff-Lag-NSI})
induce additional diagonal and non-diagonal matter potential terms
which affect neutrino flavor evolution in the Sun and in the Earth
(depending on the neutrino energy). They produce distortions in the
neutrino oscillation probabilities that can be constrained by using
solar, KamLAND, atmospheric and long-baseline neutrino data
\cite{Gonzalez-Garcia:2013usa}. Limits on all flavor non-diagonal NSI
couplings and two combinations of diagonal couplings
e.g. $\epsilon_{ii}^q-\epsilon_{\mu\mu}^q$ ($i=e,\tau$) can be derived
this way \footnote{Since the evolution of the neutrino flavor
  eigenstates is invariant under a constant shift of the Hamiltonian
  $H \rightarrow H - \mathcal{C}\mathbb{I}$ (with $\mathcal{C}$ an
  arbitrary constant and $\mathbb{I}$ the identity matrix), neutrino
  oscillation experiments can only constraint diagonal elements
  differences.}. For comparison we show the 90\% CL limits reported in
\cite{Gonzalez-Garcia:2013usa} for the large mixing angle (LMA) case
\footnote{The combined analysis of oscillation and COHERENT data
  disfavor the so-called LMA-dark solution at the $3.1\sigma$
  ($3.6\sigma$) level for up (down) quarks \cite{Coloma:2017ncl}.}  in
table \ref{tab:limits-NSI}.
\begin{table}
  \centering
  \renewcommand{\arraystretch}{1.5}
  \begin{tabular}{|c|c||c|c|}\hline
    \multicolumn{4}{|c|}{\textbf{Oscillation data (LMA)}}\\\hline
    % line 1
    $\bar\epsilon_{ee}^u$ & $[0.0,0.51]$ & $\bar\epsilon_{ee}^d$ & $[0.02,0.51]$\\\hline
    % line 2
    $\bar\epsilon_{\tau\tau}^u$ & $[-0.01,0.03]$ & $\bar\epsilon_{\tau\tau}^d$ & $[-0.01,0.03]$\\\hline
    % line 3
    $\epsilon_{e\mu}^u$ & $[-0.09,0.04]$ & $\epsilon_{e\mu}^d$ & $[-0.09,0.04]$\\\hline
    % line 4
    $\epsilon_{e\tau}^u$ & $[-0.14,0.14]$ & $\epsilon_{e\tau}^d$ & $[-0.13,0.14]$\\\hline
    % line 5
    $\epsilon_{\mu\tau}^u$ & $[-0.01,0.01]$ & $\epsilon_{\mu\tau}^d$ & $[-0.01,0.01]$\\\hline
  \end{tabular}
  \hspace{0.2cm}
  \renewcommand{\arraystretch}{1.3}
  \begin{tabular}{|c|c||c|c|}\hline
    \multicolumn{4}{|c|}{\textbf{Oscillation$+$COHERENT}}\\\hline
    % line 1
    $\epsilon_{ee}^u$ & $[0.028,0.6]$ & $\epsilon_{ee}^d$ & $[0.03,0.55]$\\\hline
    % line 2
    $\epsilon_{\mu\mu}^u$ & $[-0.088,0.37]$ & $\epsilon_{\mu\mu}^d$ & $[-0.075,0.33]$\\\hline
    % line 3
    $\epsilon_{\tau\tau}^u$ & $[-0.09,0.38]$ & $\epsilon_{\tau\tau}^d$ & $[-0.075,0.33]$\\\hline
    % line 4
    $\epsilon_{e\mu}^u$ & $[-0.073,0.044]$ & $\epsilon_{e\mu}^d$ & $[-0.07,0.04]$\\\hline
    % line 5
    $\epsilon_{e\tau}^u$ & $[-0.15,0.13]$ & $\epsilon_{e\tau}^d$ & $[-0.13,0.12]$\\\hline
    % line 6
    $\epsilon_{\mu\tau}^u$ & $[-0.01,0.009]$ & $\epsilon_{\mu\tau}^d$ & $[-0.009,0.008]$\\\hline
  \end{tabular}
  \caption{90\% CL allowed ranges for vector NSI couplings 
    for up and down quarks. Ranges in the table to the left
    are derived from neutrino oscillation data only 
    for the large mixing angle (LMA) case \cite{Gonzalez-Garcia:2013usa}, 
    while those in the table to the right arise from the combined 
    analysis of oscillation and COHERENT data \cite{Coloma:2017ncl}, and are the ones 
    we use in our analysis. In the oscillation data table $\bar\epsilon_{ii}^q$ refers to 
    $\epsilon_{ii}^q-\epsilon_{\mu\mu}^q$ ($i=e, \tau$).}
  \label{tab:limits-NSI}
\end{table}

The vector current modifies as well the C$\nu$NES cross section, and
so data from the recent measurement of the C$\nu$NES process by the
COHERENT experiment \cite{Akimov:2017ade} can be used to set bounds
on the vector NSI parameters. The low-energy neutrinos
($E_\nu\lesssim 50\,$~MeV) used by COHERENT result from
accelerator-driven protons that strike a fixed mercury target thus
producing pions. Their decay produce a monochromatic muon neutrino
flux and muons that afterward decay, thus producing anti-muon and
electron continuous neutrino fluxes. These data therefore enable
placing bounds on $\epsilon^q_{ei}$ and $\epsilon^q_{\mu i}$
($i=e, \mu, \tau$) \cite{Akimov:2017ade,Liao:2017uzy}. Overall, bounds
can be placed by two independent experimental sources (order $G_F$ and
order $G_F^2$ processes), but can be as well derived by combining
both, resulting in more competitive limits. Such an analysis has been
done in ref.\cite{Coloma:2017ncl}, and these are the limits we use in
the study of the impact of neutrino NSI in near-future DM detectors in
sec.~\ref{sec:nsi}. For reference they are displayed in table
\ref{tab:limits-NSI} (table to the right).

\section{Neutrino-nucleus elastic scattering events}
\label{sec:dm-detectors}
\subsection{Coherent neutrino-nucleus elastic cross section and NSI}
\label{sec:cnuNcs}
C$\nu$NES is a neutral-current process in which the momentum exchange
is of order $q\lesssim 100\,$~MeV. In the SM it is therefore
well-described by the effective Lagrangian
\begin{equation}
  \label{eq:eff-SM}
  \mathcal{L}_\text{Eff}^\text{SM}=-\sqrt{2}G_F
  \sum_{q=u,d}\left(\overline{\nu}_i\gamma_\mu P_L\nu_i\right)
  \overline{q}\gamma^\mu
  \left(g^q_V + g^q_A\gamma_5\right)q\ ,
\end{equation}
where $g^q_V$ and $g^q_A$ are the SM vector and axial-vector
couplings. The cross section for C$\nu$NES with spinless nuclei is
given by \cite{Freedman:1973yd,Freedman:1977xn}
\begin{equation}
  \label{eq:nu-N-x-sec-SM}
  \frac{d\sigma(E_\nu,E_r)}{dE_r}=
  \frac{G_F^2}{4\pi}\,Q^2_\text{SM}\,m_N
  \left(1-\frac{E_r}{E_r^\text{max}(E_\nu)}\right)F^2(E_r)\ .
\end{equation}
Here $m_N$ is the nucleus mass and $E_r$ the nuclear recoil energy,
whose maximum value, $E_r^\text{max}$, is given by
\begin{equation}
  \label{eq:recoil-E-max}
  E_r^\text{max}=\frac{2E_\nu^2}{m_N+2E_\nu}\, .
\end{equation}
Note that since $E_\nu\ll m_N$ ($E_\nu\lesssim 100\,$~MeV for the
neutrino-nucleus scattering process to be coherent),
$E_r/E_r^\text{max}\simeq E_rm_N/2E_\nu^2$. The atomic number and
number of neutrons are encoded in
$Q_\text{SM}=N-(1- 4\sin^2\theta_W)Z$, from which one can see the
$N^2$ enhancement of the C$\nu$NES cross section. Finally, we use the
Helm \cite{Helm:1956zz} form factor
\begin{equation}
  \label{eq:helm-f-fac}
  F^2(q\;r_n)=3\;\frac{j_1(q\;r_n)}{q\;r_n}\;e^{-q^2\,s^2/2}\ ,
\end{equation}
where $q$ can be traded to recoil energy through
$q=6.92\times 10^{-3}\sqrt{A\;E_r}\,\text{fm}^{-1}$, $j_1(x)$ is the
order-one spherical Bessel function of the first kind and for the
different parameters in (\ref{eq:helm-f-fac}) we adopt the conventions
from \cite{Lewin:1995rx}. Nuclear radius according to
\begin{equation}
  \label{eq:nuclear-radius}
  r_n^2=c^2 + \frac{7}{3}\pi^2a^2-5s^2\ ,
\end{equation}
with skin thickness given by $s=0.9\,$fm, $a=0.52\,$fm and
$c=(1.23 A^{1/3} - 0.6)\,$fm.

C$\nu$NES can take place with non-zero spin nuclei as well.  For
example, of the six most abundant Xe isotopes ($^{129}$Xe, $^{130}$Xe,
$^{131}$Xe, $^{132}$Xe, $^{134}$Xe, $^{136}$Xe) $^{129}$Xe is a
$J=1/2$ fermion state whereas $^{131}$Xe is a $J=3/2$. To properly
account for scattering with non-zero spin nuclei one has to include
contributions from the axial-vector current
in~(\ref{eq:eff-SM}). These contributions, being related with the spin
operator, are suppressed compared with those from the vector
current. Thus, to a fairly good approximation, the C$\nu$NES is
well-described by~(\ref{eq:nu-N-x-sec-SM}) regardless of the spin of
the target nuclei.

Once NSI are introduced, the C$\nu$NES cross section is no longer
flavor diagonal and so different neutrino flavors have different cross
sections. For $\nu_i-N$ it reads \cite{Barranco:2005yy,Lindner:2016wff}:
\begin{equation}
  \label{eq:cross-sec-NSI}
  \frac{d\sigma_{\nu_i}}{dE_r}=  \frac{G_F^2}{4\pi}\,Q^2_{\nu_i}\,m_N
  \left(1-\frac{E_r}{E_r^\text{max}(E_\nu)}\right)F^2(E_r)\ .
\end{equation}
where the only difference with respect to (\ref{eq:nu-N-x-sec-SM})
arises from $Q_\text{SM}\to Q_{\nu_i}$, with
\begin{equation}
  \label{eq:QNSI-i}
  Q_{\nu_i}^2=4
  \left[
    -\frac{Q_\text{SM}}{2}
    +
    N\left(\epsilon_{ii}^u + 2\epsilon_{ii}^d\right)
    +
    Z\left(2\epsilon_{ii}^u + \epsilon_{ii}^d\right)
  \right]^2
  +
  \sum_{j\neq i}
  4
  \left[
    N\left(\epsilon_{ij}^u + 2\epsilon_{ij}^d\right)
    +
    Z\left(2\epsilon_{ij}^u + \epsilon_{ij}^d\right)
  \right]^2\ .
\end{equation}
Note that for light mediators ($m_X^2\ll q^2$) this is no longer true,
and a different recoil energy dependence is expected. If this is the
case, the $\nu-N$ and WIMP-nucleon recoil spectra might differ, thus
alleviating the effects of the solar and atmospheric neutrino
backgrounds \cite{daristi:2017-c}.
\subsection{Solar neutrino flavor conversion and 
  neutrino-nucleus scattering rates}
\label{sec:sol-atm-fluxes}
Solar neutrinos ($\nu_e$) are produced in regions near the solar core
($r\lesssim 0.2 R_\odot$), with the exact distribution of neutrino
production determined by the SSM.  Once produced, electron neutrinos
are subject to flavor conversion, governed by the vacuum and matter
Hamiltonians according to
\begin{equation}
  \label{eq:evolution}
  i\frac{d}{dr}|\boldsymbol{\nu}\rangle
  =\left[
    \frac{1}{2E_\nu}\boldsymbol{U}\;
    \boldsymbol{H_\text{vac}}\;\boldsymbol{U}^\dagger
    +
    \boldsymbol{H_\text{mat}}
  \right]|\boldsymbol{\nu}\rangle\ .
\end{equation}
Here $|\boldsymbol{\nu}\rangle^T= |\nu_e, \nu_\mu, \nu_\tau \rangle^T$
refers to the neutrino flavor eigenstate basis, $r$ to the neutrino
propagation path,
$\boldsymbol{U}=\boldsymbol{U}(\theta_{23})\boldsymbol{U}(\theta_{13})\boldsymbol{U}(\theta_{12})$
is the leptonic mixing matrix ($U(\theta_{ij})$ is a $3\times 3$
rotation matrix),
$\boldsymbol{H_\text{vac}}=\text{diag}(0,\Delta m_{21}^2,\Delta
m_{31}^2)$
and in the absence of NSI
$\boldsymbol{H_\text{mat}}=\sqrt{2}G_F\,n_e(r)\text{diag}(1,0,0)$,
with $n_e(r)$ the solar electron number density. Neutrino NSI induce
additional matter potentials which change the matter Hamiltonian and
thus affect neutrino flavor evolution. In full generality the
evolution equation in (\ref{eq:evolution}) becomes
\begin{equation}
  \label{eq:ev-eq}
  i\frac{d}{dr}|\boldsymbol{\nu} \rangle
  =
  \left[
    \frac{1}{2E_\nu}\boldsymbol{U}\;
    \boldsymbol{H_\text{vac}}\;
    \boldsymbol{U}^\dagger
    +
    \sqrt{2}G_Fn_e(r)
    \sum_{f=e,u,d}
    \boldsymbol{\varepsilon^f}
  \right]|\boldsymbol{\nu} \rangle\ ,
\end{equation}
where the NSI coupling matrices $\boldsymbol{\varepsilon^f}$ read
\begin{equation}
  \label{eq:varepsilon-matrix}
   \boldsymbol{\varepsilon^f}=
   \begin{pmatrix}
     1 + \varepsilon_{ee}^f & \varepsilon_{e\mu}^f & \varepsilon_{e\tau}^f \\
     \varepsilon_{e\mu}^f & \epsilon_{\mu\mu}^f & \varepsilon_{\mu\tau}^f\\
     \varepsilon_{e\tau}^f & \varepsilon_{\mu\tau}^f  & \varepsilon_{\tau\tau}^f\\
   \end{pmatrix}\ ,
\end{equation}
with $\varepsilon_{ij}^f(r)=Y_f(r)\epsilon_{ij}^f$ ($f=e,u,d$) and
$Y_f(r)=n_f(r)/n_e(r)$. The up- and down-quark relative abundances can
be written in terms of the neutron relative abundance according to
\begin{equation}
  \label{eq:up-and-down-quark-abundances}
  Y_u = 2 + Y_n\qquad \text{and}\qquad Y_d = 1 + 2 Y_n\ ,
\end{equation}
with the neutron number density in turn calculated from the $^4$He and
$^1$H mass fractions (metallicity in the Sun amounts to less than
1.5\% so it can be safely neglected), namely
\begin{equation}
  \label{eq:neutron-density}
  n_n(r)\simeq\frac{X(^4\text{He})}{2X(^1\text{H})+X(^4\text{He})}\ .
\end{equation}
Since we are interested in C$\nu$NES, from now on we set
$\epsilon^e_{ij}=0$ and following
refs. \cite{Gonzalez-Garcia:2013usa,Coloma:2017ncl} we do not consider
simultaneous contributions from up- and down-quark couplings. For all
the SSM related quantities we use those from the BS05 SSM
\cite{Bahcall:2004pz}.

To a fairly good approximation, flavor conversion probabilities can be
calculated in the mass dominance limit, $\Delta m_{31}^2\to \infty$.
In this limit, and with vanishing CP-violating phases, neutrino
propagation takes place in the basis
$|\boldsymbol{\tilde
  \nu}\rangle=\boldsymbol{U_{13}}^T\boldsymbol{U_{23}}^T|\boldsymbol{\nu}\rangle$,
which neglecting terms proportional to $\sin\theta_{13}$ corresponds
to the following neutrino eigenstates:
\begin{equation}
  \label{eq:neutrino-propagation-states-two-fl}
  \tilde\nu_e\simeq \nu_e\ ,\qquad
  \tilde\nu_\mu= \cos\theta_{23}\nu_\mu - \sin\theta_{23}\nu_\tau\ ,\qquad
  \tilde\nu_\tau\simeq \sin\theta_{23}\nu_\mu + \cos\theta_{23}\nu_\tau\ .
\end{equation}
In this basis, which we dub as \textit{propagation basis},
$\tilde\nu_e-\tilde \nu_\tau$ and $\tilde\nu_\mu-\tilde \nu_\tau$
mixing is of order
$\xi_{ij}=G_Fn_q(r)\epsilon_{ij}^q/(\Delta m_{31}^2/E_\nu)$.  Taking
$n_q\simeq 10^{26}\,\text{cm}^{-3}$ (up-quark number density at about
$r=0.1 R_\odot$), $E_\nu\lesssim 20\,$MeV and
$\Delta m_{31}^2=2.55\times 10^{-3}\,\text{eV}^2$ (BFPV for normal
ordering) \cite{deSalas:2017kay}, one can see that even for order-one
NSI couplings $\xi_{ij}\ll 1$.  This means that in the mass dominance
limit, $\tilde\nu_\tau$ is decoupled and so neutrino flavor evolution
can be studied in the two-flavor approximation with electron neutrino
flavor conversion determined by
$1-\langle\mathcal{P}_{ee}(E_\nu)\rangle$. Thus, in what follows we
discuss the calculation of the averaged survival probability
$\langle\mathcal{P}_{ee}(E_\nu)\rangle$. First of all, due to the
quark and electron densities $\mathcal{P}_{ee}$ depends not only on
$E_\nu$ but also on the neutrino propagation path
$r$. $\mathcal{P}_{ee}(E_\nu,r)$ can be written as \cite{Kuo:1986sk}
\begin{equation}
  \label{eq:survival-prob}
  \mathcal{P}_{ee}(E_\nu,r)=\cos^4\theta_{13}
  \,\mathcal{P}_\text{eff}(E_\nu,r)
  +
  \sin^4\theta_{13}\ ,
\end{equation}
where the $r$ dependence is introduced by the effective probability
given by \cite{Parke:1986jy} \footnote{We have checked that
  $\gamma^{-1}\ll 1$, which guarantees neutrino adiabatic propagation,
  and so have taken the level-crossing probability $P_c\to 0$.}
\begin{equation}
  \label{eq:eff-prob}
  \mathcal{P}_\text{eff}(E_\nu,r)=
  \frac{1+\cos2\theta_M(r)\cos2\theta_{12}}{2}\ .
\end{equation}
The mixing angle in matter is calculated from the diagonalization of
the $2\times 2$ Hamiltonian
\begin{equation}
  \label{eq:2times2-ham}
  \boldsymbol{H}=\frac{1}{4E_\nu}
  \begin{pmatrix}
    -\Delta m_{21}^2 \cos2\theta_{12} + A 
    & \Delta m_{21}^2 \sin2\theta_{12} + B\\
    \Delta m_{21}^2 \sin2\theta_{12} + B 
    & \Delta m_{21}^2 \cos2\theta_{12} - A
  \end{pmatrix}\ ,
\end{equation}
where
\begin{align}
  \label{eq:A-B}
  A=4\sqrt{2}E_\nu G_F n_e(r) 
  \left[\frac{\cos^2\theta_{13}}{2} - Y_q(r)\varepsilon_D\right]\ ,\qquad
  B=4\sqrt{2}E_\nu G_F n_e(r) Y_q(r)\varepsilon_N \ .
\end{align}
Note that in the limit $\epsilon_{ij}^q=0$ and $\cos\theta_{13}=0$,
$A$ reduces to the SM term and $B$ vanishes. The parameters
$\varepsilon_D$ and $\varepsilon_N$ result from the rotation from the
flavor to the \textit{propagation basis} and read
\cite{Gonzalez-Garcia:2013usa}:
\begin{align}
  \label{eq:epsilonD}
  \varepsilon_D&=
  -\frac{c_{13}^2}{2}\epsilon_{ee}^q
  +\frac{\left[c_{13}^2 - \left(s_{23}^2 - s_{13}^2 c_{23}^2\right)\right]}{2}
  \epsilon_{\mu\mu}^q
  +\frac{\left(s_{23}^2 - c_{23}^2s_{13}^2\right)}{2}
  \epsilon_{\tau\tau}^q + s_{13}c_{13}s_{23}\epsilon_{e\mu}^q
  \nonumber\\
  &\;\quad
  + s_{13}c_{13}c_{23}\epsilon_{e\mu}^q
  - c_{23}s_{23}\epsilon_{\mu\tau}^q\ ,
  \\
  \label{eq:epsilonN}
  \varepsilon_N&= 
  -s_{13}c_{23}s_{23}\epsilon_{\mu\mu}^q
  + s_{13}c_{23}s_{23}\epsilon_{\tau\tau}^q
  + c_{13}c_{23}\epsilon_{e\mu}^q
  - c_{13}s_{23}\epsilon_{e\tau}^q
  + s_{13}\left(s_{23}^2 - c_{23}^2\right)\epsilon_{\mu\tau}^q\ .
\end{align}
\begin{figure}
  \centering
  \includegraphics[scale=0.75]{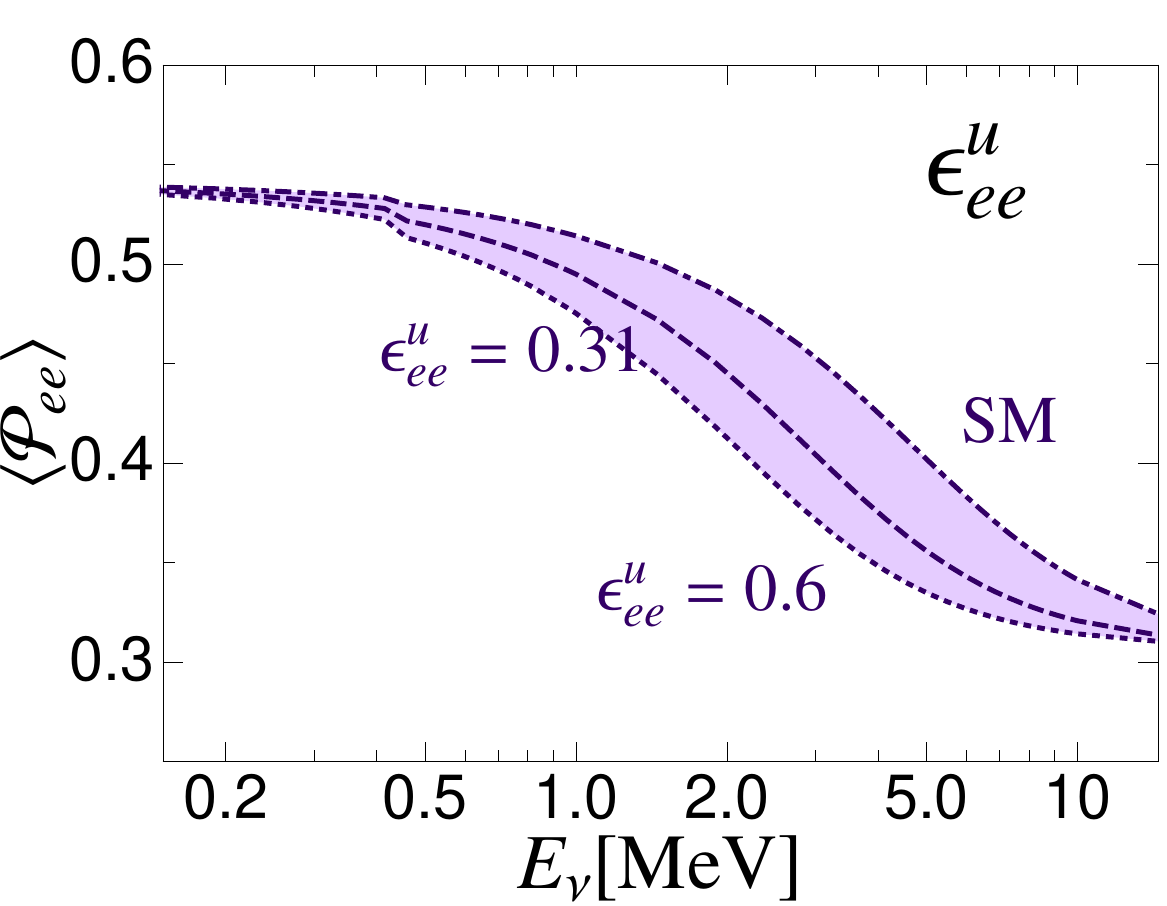}
  \includegraphics[scale=0.75]{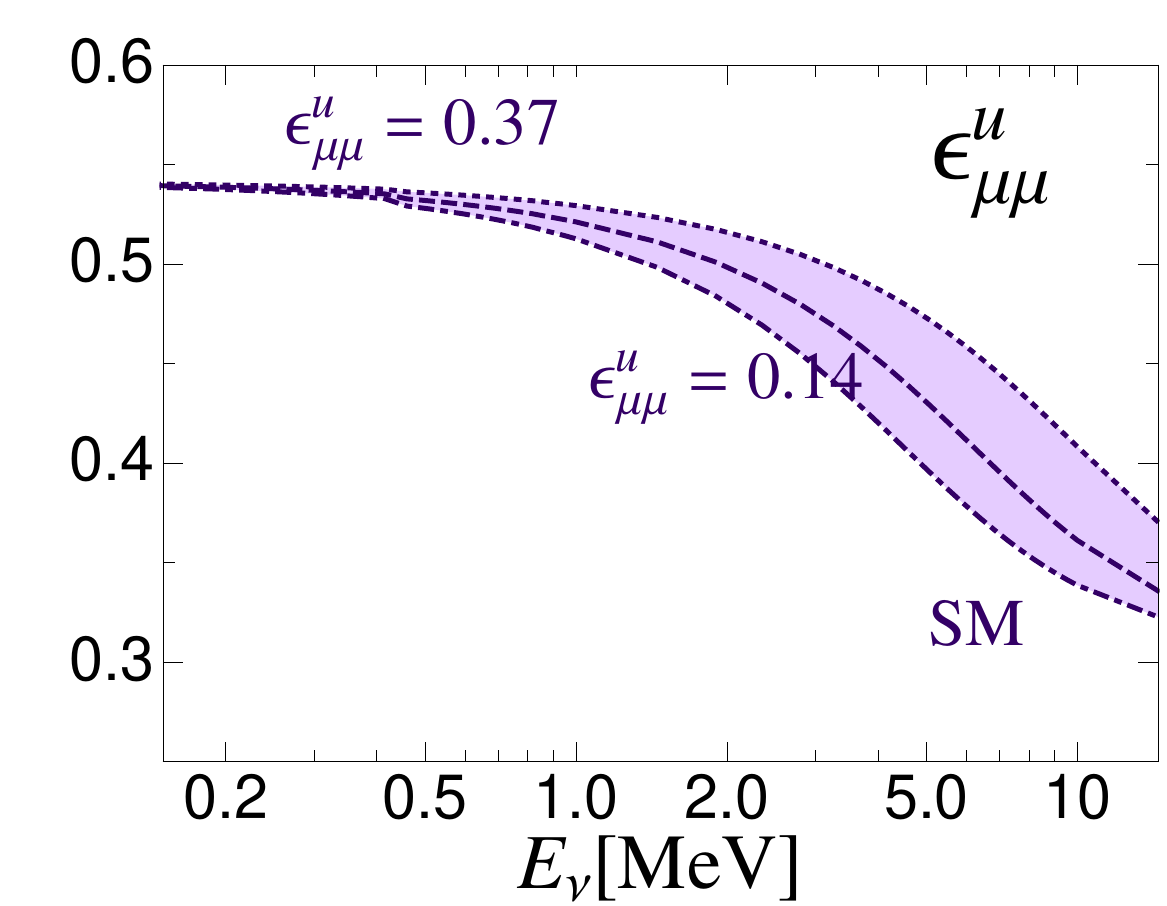}
  \caption{Averaged solar neutrino flavor oscillation probabilities
    for $\epsilon_{ee}^u\neq 0$ (left-hand side) and
    $\epsilon_{\mu\mu}^u\neq 0$ (right-hand side) as a function of the
    neutrino energy. Both NSI parameters have been varied in the
    ranges allowed by COHERENT and oscillation data
    \cite{Coloma:2017ncl} (see tab. \ref{tab:limits-NSI},
    Oscillation+COHERENT). In both plots, the dotted-dashed curve
    refers to the SM case, the dotted curve to the amplitude obtained
    with the largest value for the corresponding coupling and the
    dashed for an intermediate value. The remaining NSI couplings in
    both cases have been put to zero.}
  \label{fig:prob}
\end{figure}
Here we have adopted the notation $\cos\theta_{ij}=c_{ij}$ and
$\sin\theta_{ij}=s_{ij}$. From eq. (\ref{eq:2times2-ham}) the mixing
angle in matter can be straightforwardly calculated:
\begin{equation}
  \label{eq:mixing-angle}
  \cos2\theta_M(r) = \frac{\Delta m_{12}^2\cos2\theta_{12}-A}
  {\sqrt{\left(\Delta m_{12}^2\cos2\theta_{12}-A\right)^2
      +
  \left(\Delta m_{12}^2\sin2\theta_{12}+B\right)^2}}\ .
\end{equation}
Eqs. (\ref{eq:survival-prob}) and (\ref{eq:eff-prob}) combined with
eqs. (\ref{eq:A-B})-(\ref{eq:mixing-angle}) allow the determination of
$\mathcal{P}_{ee}(E_\nu,r)$ in terms of neutrino oscillation
parameters, electron and quark number densities and NSI
parameters. The averaged survival probability is then obtained by
integrating over $r$ taking into account the distribution of neutrino
production in the Sun \cite{Gonzalez-Garcia:2013usa}:
\begin{equation}
  \label{eq:averaged-survival-prob}
  \langle\mathcal{P}_{ee}(E_\nu)  \rangle=
  \frac{\sum_\alpha\Phi_\alpha(E_\nu)\int_0^1
    \,dr\rho_\alpha(r)\,\mathcal{P}_{ee}(E_\nu,r)}
  {\sum_\alpha\Phi_\alpha(E_\nu)}\ ,
\end{equation}
where $\Phi_\alpha(E_\nu)$ stands for neutrino fluxes and
$\rho_\alpha(r)$ for the distribution of neutrino production,
$\alpha=pp,^7\text{Be}\,\text{(ground and excited state)}, pep, hep,
^8\text{B}, ^{13}\text{N}, ^{15}\text{O}, ^{17}\text{F}$.
In practice, to calculate $\langle\mathcal{P}_{ee}(E_\nu) \rangle$ we
have followed a single-parameter analysis, varying only one NSI
coupling at a time in the ranges allowed by the constraints derived
from neutrino oscillation and COHERENT data
\cite{Coloma:2017ncl}. These ranges have been split according to
$\epsilon_{ij}^q|_{n+1}=\epsilon_{ij}^q|_{n}+\delta\epsilon_{ij}^q$
with
$\delta\epsilon_{ij}^q=
\left(\epsilon_{ij}^q|_\text{max}-\epsilon_{ij}^q|_\text{min}\right)/10$.
And the neutrino energy, $E_\nu$, has been varied from 0.145 MeV to
14.5 MeV. The neutrino oscillation parameters have been fixed to their
BFPVs according to ref.\cite{deSalas:2017kay}. For each NSI parameter
we thus have generated 11 data sets (11 averaged survival
probabilities), including the SM case ($\epsilon_{ij}^q=0$).

Focusing on up-quark NSI couplings (results for down-quark parameters
resemble those of the up-quark case), the averaged survival
probability deviates the most from the SM expectation for
flavor-diagonal NSI and $\epsilon_{e\tau}^u$, being more pronounced
for the former than for the latter. Fig. \ref{fig:prob} shows the
result for $\langle\mathcal{P}_{ee}\rangle$ for the cases
$\epsilon_{ee}^u\neq 0$ and $\epsilon_{\mu\mu}^u\neq 0$ (results for
$\epsilon_{\tau\tau}^u\neq 0$ are pretty close to those found for
$\epsilon_{\mu\mu}^u\neq 0$).

It can be seen that the range of variation of
$\langle\mathcal{P}_{ee}\rangle$ has about the same amplitude in both
cases. However its behavior with increasing values of the NSI
couplings is rather different. For $\epsilon_{ee}^u\neq 0$, the
averaged survival probability decreases with increasing values of
$\epsilon_{ee}^u$ and its largest value (as a function of $E_\nu$) is
indeed found when $\epsilon_{ee}=0$. For $\epsilon_{\mu\mu}^u\neq 0$,
it is the other way around. Increasing values of the parameter leads
to larger $\langle\mathcal{P}_{ee}\rangle$. This behavior can be
readily understood as follows. In the SM case limit,
$\epsilon_{ij}^u=0$, eq. (\ref{eq:mixing-angle}) reduces to
\begin{equation}
  \label{eq:mix-limit-epsijEq0}
  \cos2\theta_M=\frac{\Delta m_{21}^2\cos2\theta_{12}
    - 2\sqrt{2}E_\nu G_Fn_e\,c_{13}^2}
  {\sqrt{\left(\Delta m_{21}^2\cos2\theta_{12}
    - 2\sqrt{2}E_\nu G_Fn_e\,c_{13}^2\right)^2
  +\Delta m_{21}^4\sin^22\theta_{12}}}\ .
\end{equation}
The presence of the $\epsilon_{ee}^u$ and $\epsilon_{\mu\mu}^u$
couplings changes this equation by shifting either the first term in
the denominator (numerator) or both. As can be seen from
eqs. (\ref{eq:A-B})-(\ref{eq:mixing-angle}) the shift in the numerator
of eq. (\ref{eq:mix-limit-epsijEq0}) due to $\epsilon_{ee}^u$ is
always negative. Thus, as $\epsilon_{ee}^u$ increases $\cos2\theta_M$
decreases and accordingly $\langle\mathcal{P}_{ee}\rangle$ decreases
(see eqs. (\ref{eq:survival-prob}) and (\ref{eq:eff-prob})). The shift
due to $\epsilon_{\mu\mu}^u$ is instead positive thus leading to
larger values of $\cos2\theta_M$ as $\epsilon_{\mu\mu}^u$ increases.

\begin{figure}
  \centering
  \includegraphics[scale=0.75]{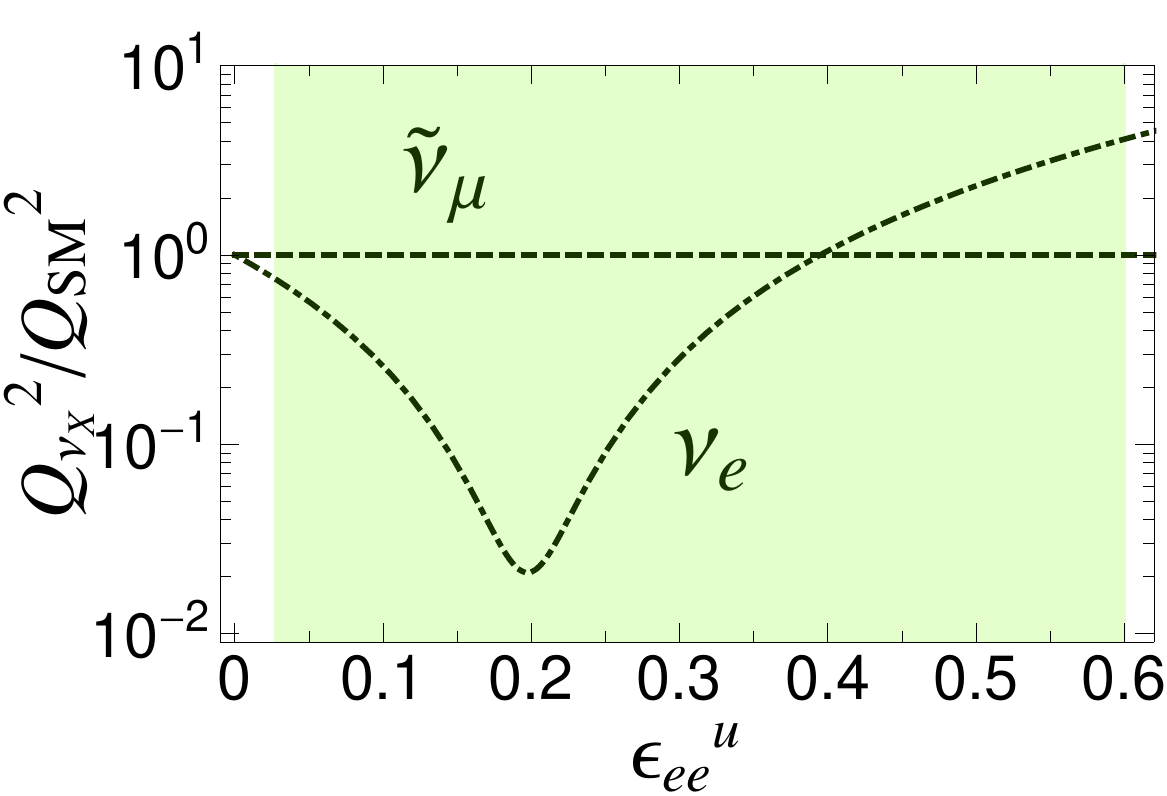}
  \includegraphics[scale=0.75]{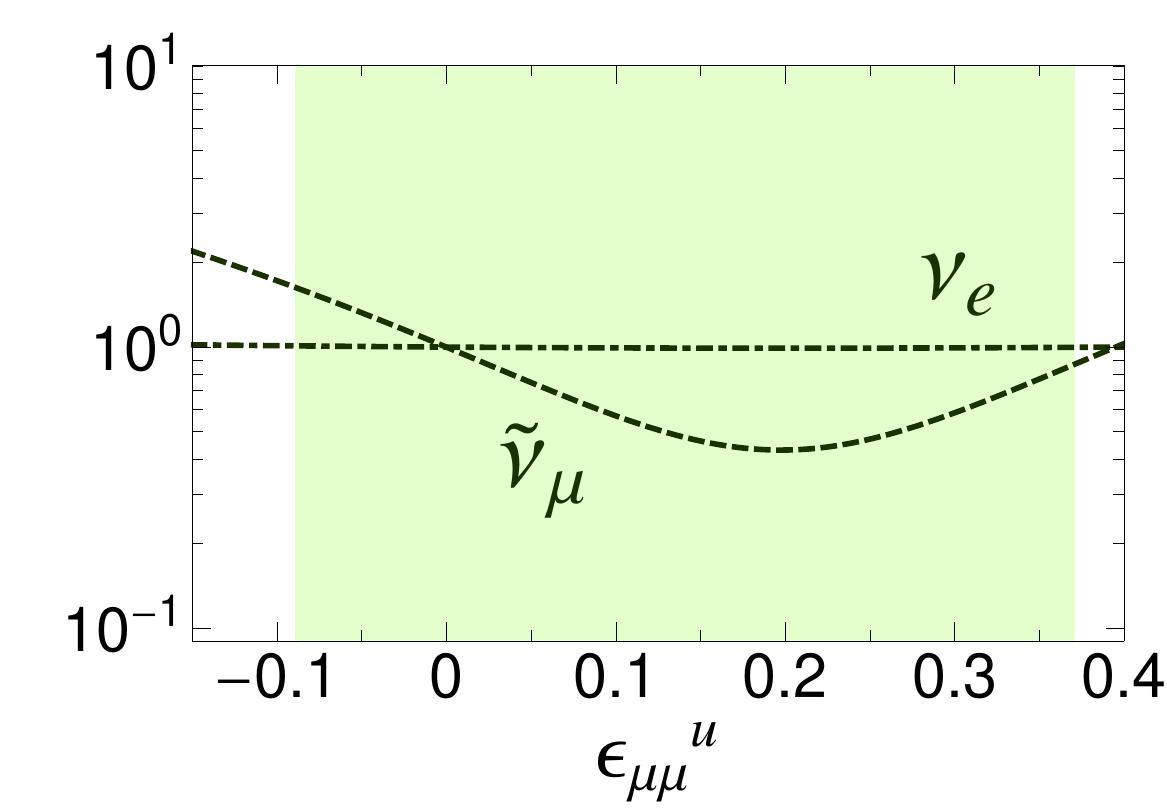}
  \caption{C$\nu$NES cross sections normalized to the SM value as a
    function of the NSI parameters $\epsilon_{ee}^u$ (left-hand graph)
    and $\epsilon_{\mu\mu}^u$ (right-hand graph). Solar neutrino
    oscillations are analyzed in the mass dominance limit and so
    $\tilde \nu_\mu\simeq c_{23}\nu_\mu - s_{23}\nu_\tau$. The shaded
    regions indicate current allowed ranges for both parameters as
    required by neutrino oscillation and COHERENT data
    \cite{Coloma:2017ncl}.}
  \label{fig:cross-sections-solar}
\end{figure}

As we have pointed out, in the mass dominance limit solar neutrino
flavor conversion can be reduced to a two-flavor oscillation problem,
in which a mainly $|\nu_e\rangle$ state oscillates to a
$|\tilde \nu_\mu\rangle$ state which is a superposition of
$|\nu_\mu\rangle$ and $|\nu_\tau\rangle$ flavor eigenstates. Thus, in
an Earth-based detector, neutrino-nucleus scattering will take place
with either $|\nu_e\rangle$ or $|\tilde \nu_\mu\rangle$ and so the
calculation of neutrino-nucleus scattering rates has to be done in the
propagation basis as well. Rotating the effective Lagrangian
in (\ref{eq:eff-Lag-NSI}) to this basis, after dropping the
axial-vector coupling matrix, results in
\begin{equation}
  \label{eq:eff-Lag-propagation-matrix}
  \mathcal{L}_\text{Eff}=-\sqrt{2}G_F\sum_{q=u,d}\overline{\boldsymbol{\tilde\nu}}
  \gamma_\mu P_L\boldsymbol{\widetilde{\epsilon^{q}}} \;\boldsymbol{\tilde \nu}
  \;\;\overline{q}\gamma^\mu q\ ,
\end{equation}
with the NSI couplings in the propagation basis related with the NSI
couplings in the flavor basis through the following transformation
\begin{equation}
  \label{eq:prop-to-flav-couplings}
  \boldsymbol{\widetilde{\epsilon^q}}=
  \boldsymbol{U}(\theta_{13})^T\;\boldsymbol{U}(\theta_{23})^T\,
  \boldsymbol{\epsilon^q}\,\boldsymbol{U}(\theta_{23})\;\boldsymbol{U}(\theta_{13})\ .
\end{equation}
The differential cross sections for $\tilde\nu_e-N\simeq \nu_e-N$ and
$\tilde\nu_\mu-N$ can then be obtained from (\ref{eq:cross-sec-NSI})
and (\ref{eq:QNSI-i}) by $\epsilon_{ij}^q\to \tilde\epsilon_{ij}^q$.
Results for the cases $\epsilon_{ee}^u\neq 0$ and
$\epsilon_{\mu\mu}^u\neq 0$ are shown for both neutrino flavors in
fig. \ref{fig:cross-sections-solar}. As can be seen, deviations on
either $d\sigma_{\nu_e}/dE_r$ or $d\sigma_{\tilde\nu_\mu}/dE_r$ are
possible. The presence of NSI can enhance but also can deplete the
flavored cross sections, thus potentially leading to substantial
deviations of the expected neutrino-nucleus scattering event rate.

Assuming a 100\% efficiency and with the oscillation probabilities and
the $\nu-N$ differential cross sections calculated in the propagation
basis, the $\nu-N$ event rate as a function of the recoil energy $E_r$
can be written according to \cite{Kuo:1989qe}:
\begin{equation}
  \label{eq:event-rate-solar}
  \frac{dR_\nu^\odot}{dE_r}=N_T\,\sum_\alpha\int_{E_\text{min}}^\infty\,
  \Phi_\alpha(E_\nu)
  \left[
   \langle\mathcal{P}_{ee}\rangle
    \frac{d\sigma_{\nu_e}}{dE_r}
    +
     \left(1- \langle\mathcal{P}_{ee}\rangle\right)
    \frac{d\sigma_{\tilde\nu_\mu}}{dE_r}
  \right]\,dE_\nu\ .
\end{equation}
Here $N_T=N_A/A$ ($N_A=6.022\times 10^{29}\,\text{ton}^{-1}$),
$E_r^\text{min}=\sqrt{m_N\,E_r/2}$ and we include neutrinos emitted in
the $pp$ and CNO chains, $\alpha=$$pp$,$^7\text{Be}\,\text{(ground
  and excited state)}$, $pep$, $hep$, $^8\text{B}$,
$^{13}\text{N}$,
$^{15}\text{O}$,
$^{17}\text{F}$,
as in (\ref{eq:averaged-survival-prob}). The neutrino fluxes are in
units of $\text{cm}^{-2}\,\text{year}^{-1}\,\text{keV}^{-1}$,
so the differential rates are in units of
$\text{ton}^{-1}\,\text{year}^{-1}\,\text{keV}^{-1}$
and the total event rates, $R_\nu^\odot$,
are then measured in $\text{ton}^{-1}\,\text{year}^{-1}$.
The latter are calculated from~(\ref{eq:event-rate-solar}) by
integration over $E_r$
from $E_r^\text{th}$
to 100 keV, with $E_r^\text{th}\subset
[10^{-3},10^2]\,$~keV and varied in steps
$\delta E_r=(E_r^{n+1}-E_r^{n})/50$.
\subsection{Atmospheric neutrino-nucleus scattering rates}
\label{sec:sol-atm-fluxes}
In contrast to solar neutrinos, matter effects for sub-GeV atmospheric
neutrinos are negligible and flavor conversion can be well described
by vacuum oscillations \cite{Friedland:2004ah}. This can be readily
seen by considering the PREM model \cite{Dziewonski:1981xy}, for which
$\langle
n_e^{\oplus}\rangle=3.25\times 10^{24}\,\text{cm}^{-3}$ and $\langle
Y_n^{\oplus}\rangle=1.137$ both in the Earth core. With these values,
using the BFPV for $\Delta
m_{31}^2$ \cite{deSalas:2017kay} and taking
$E_\nu=100\,$~MeV
so to minimize the vacuum term contribution one gets (see
eq. (\ref{eq:up-and-down-quark-abundances}))
\begin{equation}
  \label{eq:vacuum-to-matter-ratio}
  \frac{H_\text{mat}}{H_\text{vac}}\simeq
  0.2\times 
  \left(\frac{2.55\times 10^{-3}}{\Delta m_{31}^2}\right)
  \left(\frac{E_\nu}{100\,\text{MeV}}\right)
  \left(\frac{n^{\oplus}_e}{3.25\times 10^{24}\,\text{cm}^{-3}}\right)
  \left(\frac{Y_u}{3.137}\right)\,\epsilon_{ij}^u\ .
\end{equation}
Note that this is the largest contribution from $H_\text{mat}$
to the neutrino flavor evolution equation, thus implying that matter
effects for sub-GeV neutrinos can be fairly ignored.

\begin{figure}
  \centering
  \includegraphics[scale=0.75]{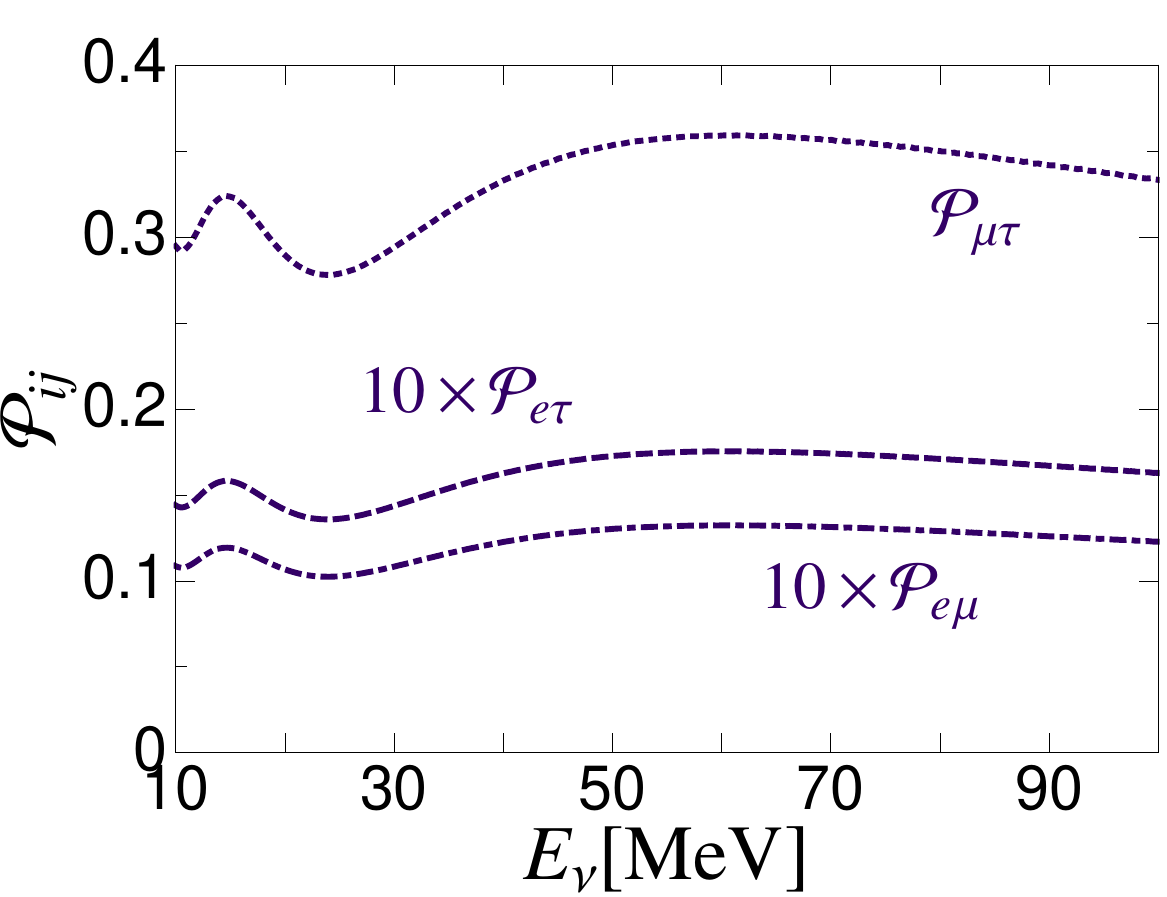}
  \caption{Atmospheric neutrino oscillations for sub-GeV neutrinos as
    a function of the neutrino energy $E_\nu$.
    For the calculation $\Delta
    m_{31}^2$ has been fixed to its BFPV according to
    ref. \cite{deSalas:2017kay}.}
  \label{fig:atmospheric-prob}
\end{figure}
Since only vacuum oscillations are relevant, the full three-flavor treatment
is straightforward. In this case the oscillation probabilities can be
further simplified by considering the limit $\Delta
m_{21}^2\to 0$. They read \cite{Akhmedov:1999uz}
\begin{equation}
  \label{eq:atmospheric-oscillations}
  \mathcal{P}_{ij}(E_\nu,L)=4\left|U_{i3}\right|^2\left|U_{j3}\right|^2
  \sin^2\left(\frac{\Delta m_{31}^2}{4E_\nu}\,L\right)\ ,
\end{equation}
where $L$
refers to the distance covered by a neutrino from production at a height
$h$ (perpendicular to the surface) to detection and is given by:
\begin{equation}
  \label{eq:L-atm}
  L=\sqrt{(R_\oplus + h)^2-R_\oplus^2\sin^2\theta_z}
  -
  R_\oplus\cos\theta_z\ ,
\end{equation}
with $\theta_z$
the zenith angle of the neutrino trajectory. The oscillation
probability in (\ref{eq:atmospheric-oscillations}) is therefore a
function of production height $h$
and zenith angle $\theta_z$.
The distribution of vertical production height for sub-GeV neutrinos
was calculated in \cite{Battistoni:2005pd} and its mean was found to
be $\langle
h\rangle\simeq
18\,$~km. Thus, rather than using the distribution we fix
$h$
to this value and integrate over the zenith angle. The oscillation
probabilities relevant for the determination of the $\nu-N$
event rate can then be written as
\begin{equation}
  \label{eq:prob-atm-only-energy-dependent}
  \mathcal{P}_{ij}(E_\nu)=\frac{1}{\pi}\int_{-1}^1\,d\cos\theta\,
  \mathcal{P}_{ij}(E_\nu,L)\ .
\end{equation}
As expected, neutrino flavor oscillations in this case are only
relevant for $\nu_\mu-\nu_\tau$
conversion. The result is shown in fig. \ref{fig:atmospheric-prob}.

\begin{figure}
  \centering
  \includegraphics[scale=0.75]{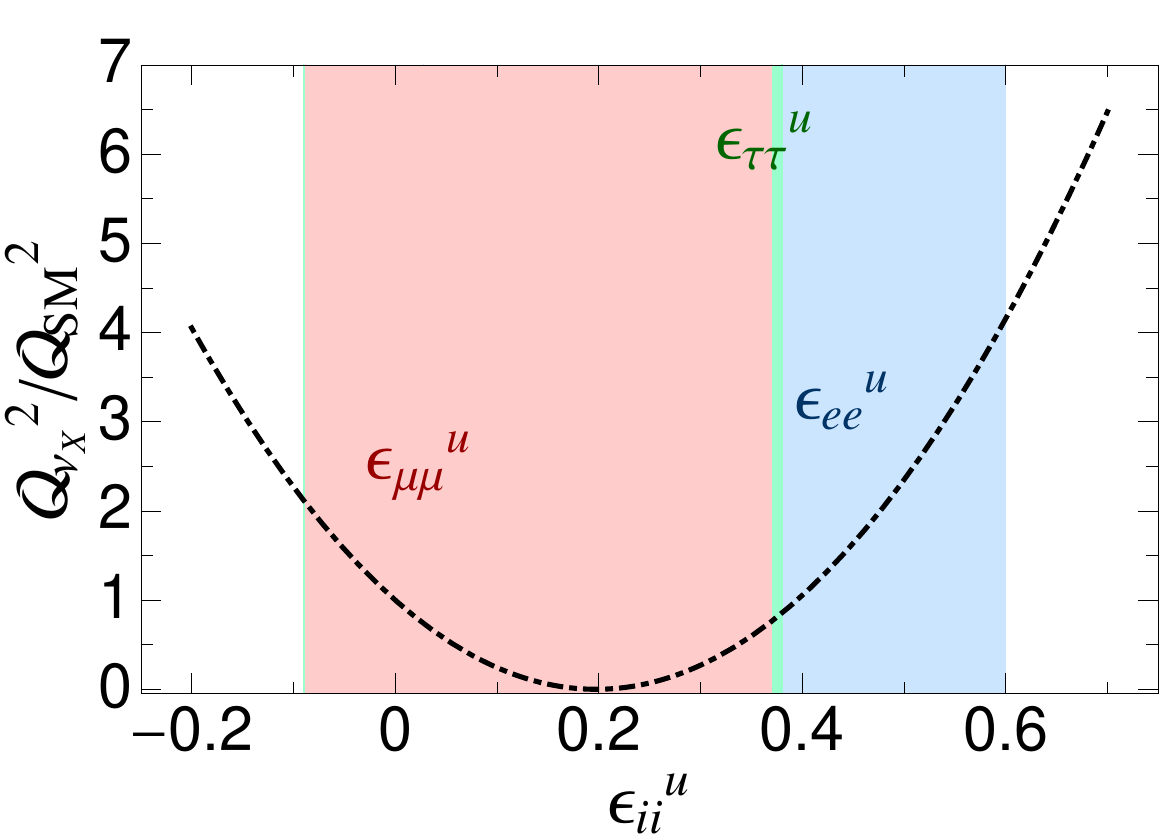}
  \caption{C$\nu$NES cross section normalized to the SM value as a
    function of diagonal NSI couplings. Shaded regions refer to
    current limits on $\epsilon_{ii}^u$ ($i=e,\mu,\tau$) as required
    by neutrino oscillation and COHERENT data
    \cite{Coloma:2017ncl}. The blue shaded region ($\epsilon_{ee}^u$) covers values down to
    0.028 (see tab. \ref{tab:limits-NSI}). These are the cross
    sections which deviate the most from the SM value and that
    therefore lead to the largest deviations on the number of
    neutrino-nucleus scattering events in the atmospheric sector.}
  \label{fig:atm-xsec}
\end{figure}

Since we set CP-violating phases to zero, the differential event rates
associated with $\nu_e+\bar\nu_e$ and $\nu_\mu+\bar\nu_\mu$ fluxes can
be written according to
\begin{eqnarray}
  \label{eq:atm-event-rates-elec}
  \frac{dR_{\nu_e+\bar\nu_e}^\text{Atm}}{dE_r}&=&N_T\int_{E_\text{min}}^\infty
  \Phi_{\nu_e+\bar\nu_e}(E_\nu)
  \left[
  \left(1-\mathcal{P}_{e\mu}-\mathcal{P}_{e\tau}\right)\frac{d\sigma_{\nu_e}}{dE_r}
  +
  \mathcal{P}_{e\mu}\frac{d\sigma_{\nu_\mu}}{dE_r}
  +
  \mathcal{P}_{e\tau}\frac{d\sigma_{\nu_\tau}}{dE_r}
  \right]\ ,
  \\
  \label{eq:atm-event-rates-muon}
  \frac{dR_{\nu_\mu+\bar\nu_\mu}^\text{Atm}}{dE_r}&=&N_T\int_{E_\text{min}}^\infty
  \Phi_{\nu_\mu+\bar\nu_\mu}(E_\nu)
  \left[
  \mathcal{P}_{e\mu}\frac{d\sigma_{\nu_e}}{dE_r}
  +
  \left(1-\mathcal{P}_{e\mu}-\mathcal{P}_{\mu\tau}\right)
  \frac{d\sigma_{\nu_\mu}}{dE_r}
  +
  \mathcal{P}_{\mu\tau}\frac{d\sigma_{\nu_\tau}}{dE_r}
  \right]\ .
\end{eqnarray}
In (\ref{eq:atm-event-rates-muon}) we assume time reversal
invariance. We use the atmospheric neutrino fluxes from
ref. \cite{Battistoni:2005pd}, derived with the FLUKA Monte Carlo
simulation package \cite{Ferrari:2005zk} and which provides fluxes for
$\nu_e$, $\nu_\mu$ and the corresponding anti-neutrinos. The full
differential $\nu-N$ event rate is obtained from
(\ref{eq:atm-event-rates-elec}) and (\ref{eq:atm-event-rates-muon})
and the total event rate as in the solar case, integrating over $E_r$
from $E_r^\text{th}$ to 100 keV, with
$E_r^\text{th}\subset [10^{-3},10^2]\,$~keV and varied in steps
$\delta E_r=(E_r^{n+1}-E_r^{n})/50$. The flavored cross sections are
then given by~(\ref{eq:cross-sec-NSI}) for which one finds sizable
deviations from the SM only for diagonal NSI couplings, as expected
given the tight constraints on the non-diagonal, particularly
$\epsilon_{e\mu}^u$ and $\epsilon_{\mu\tau}^u$. The result for
diagonal couplings is shown in fig.~\ref{fig:atm-xsec}, in which all
couplings but one have been assumed to be zero.
\section{NSI at ton-size DM Xe detectors}
\label{sec:nsi-dm-detectors}
\begin{figure}[t]
  \centering
  \includegraphics[scale=0.45]{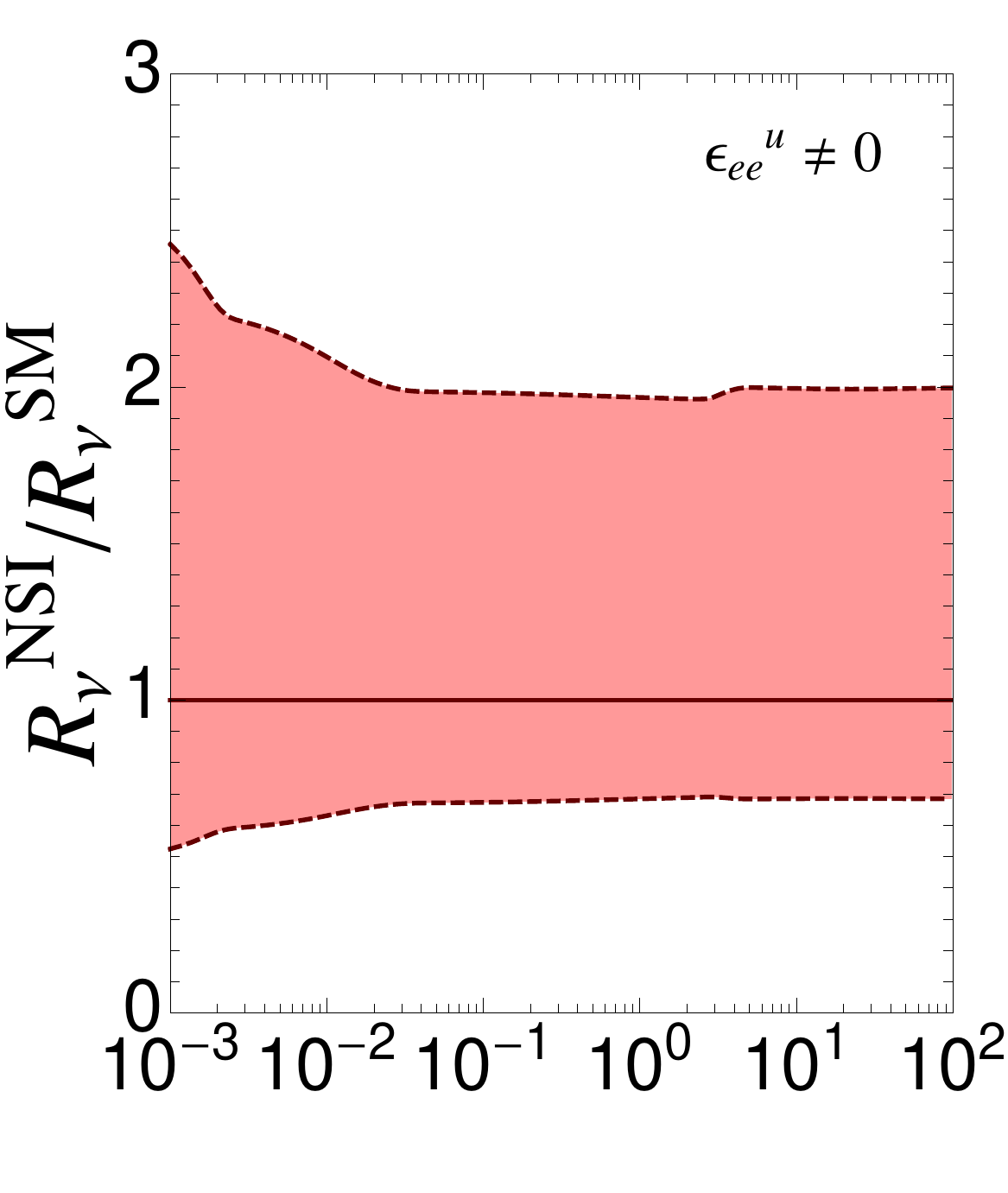}
  \includegraphics[scale=0.45]{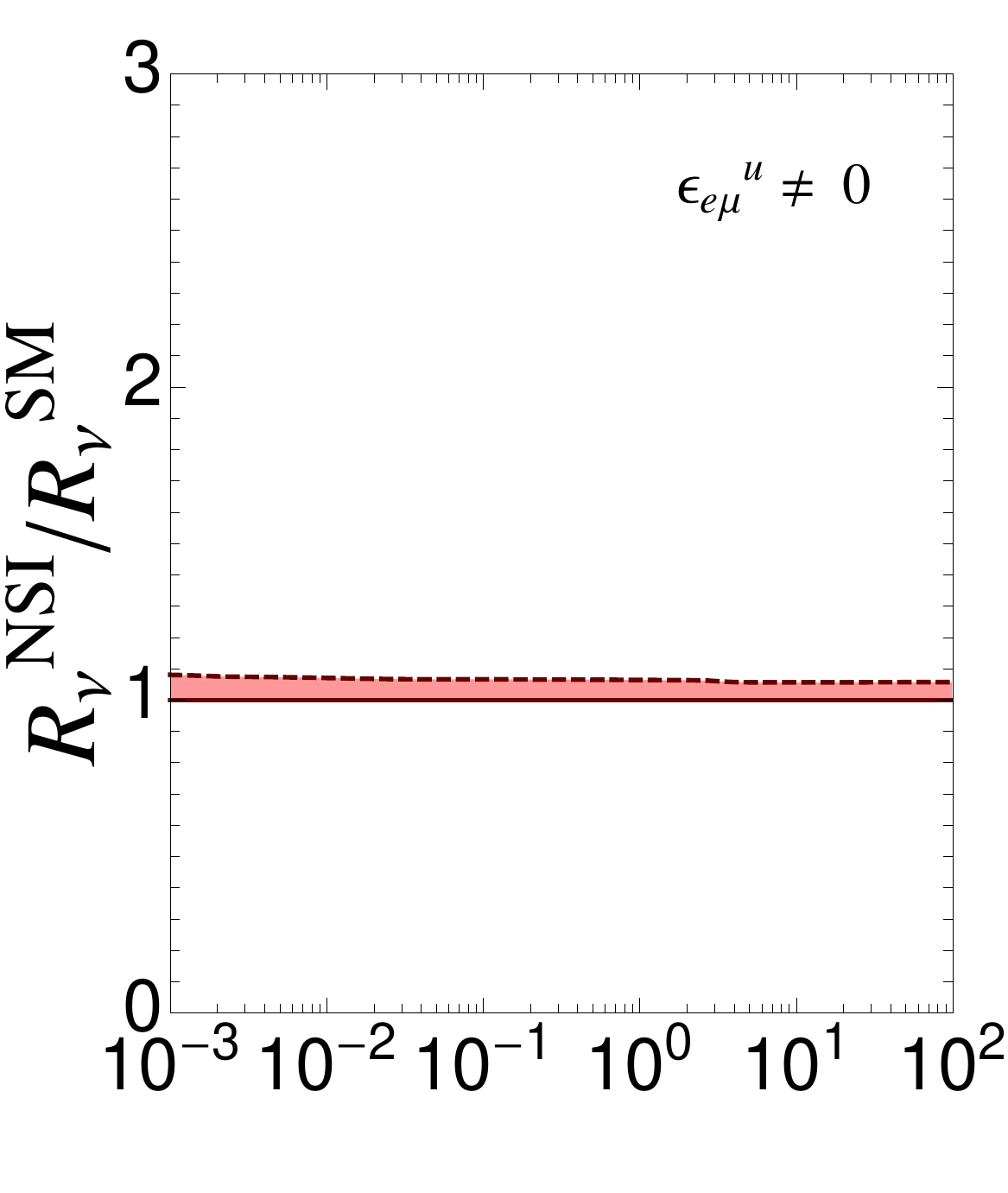}
  \includegraphics[scale=0.45]{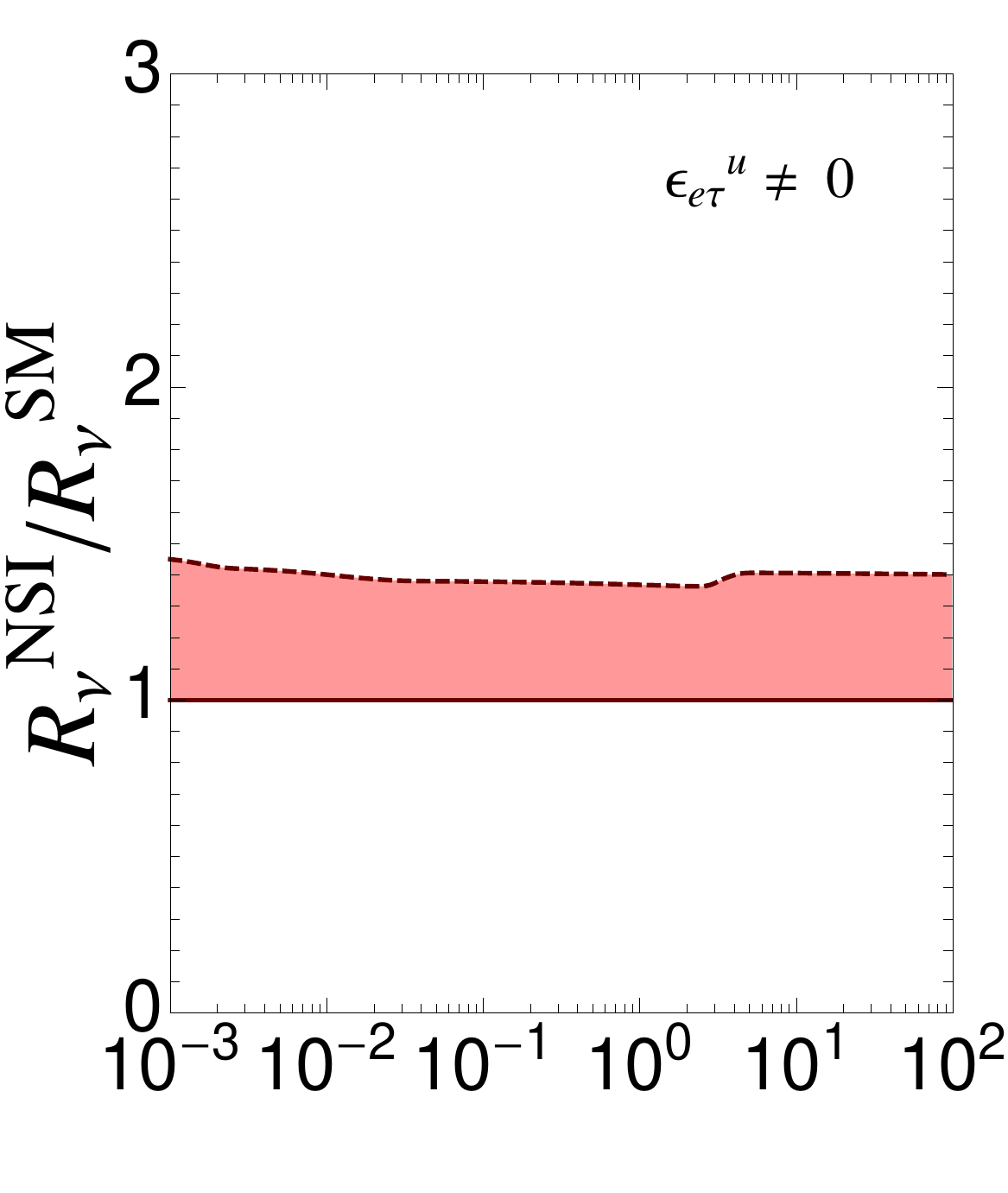}
  \includegraphics[scale=0.45]{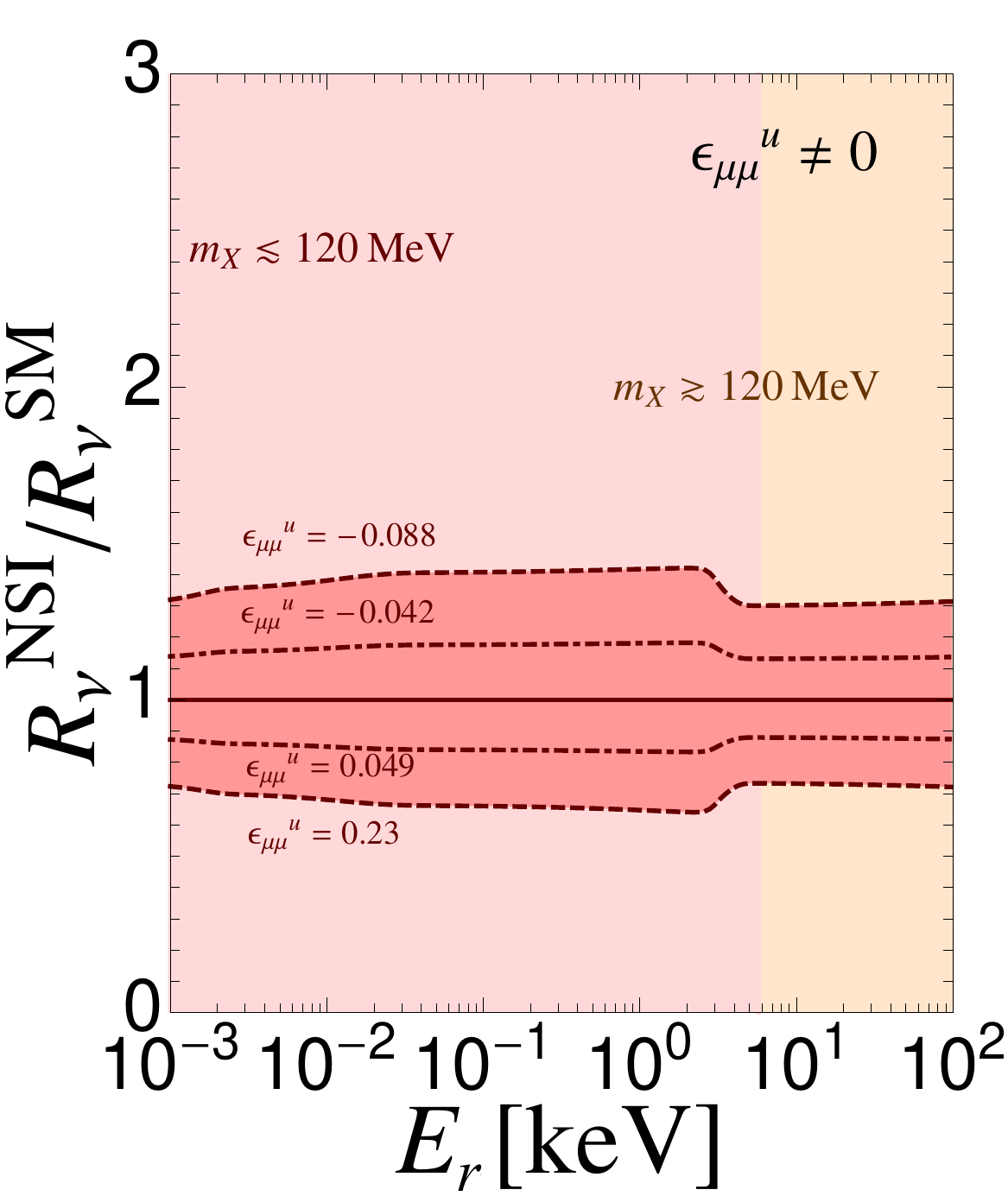}
  \includegraphics[scale=0.45]{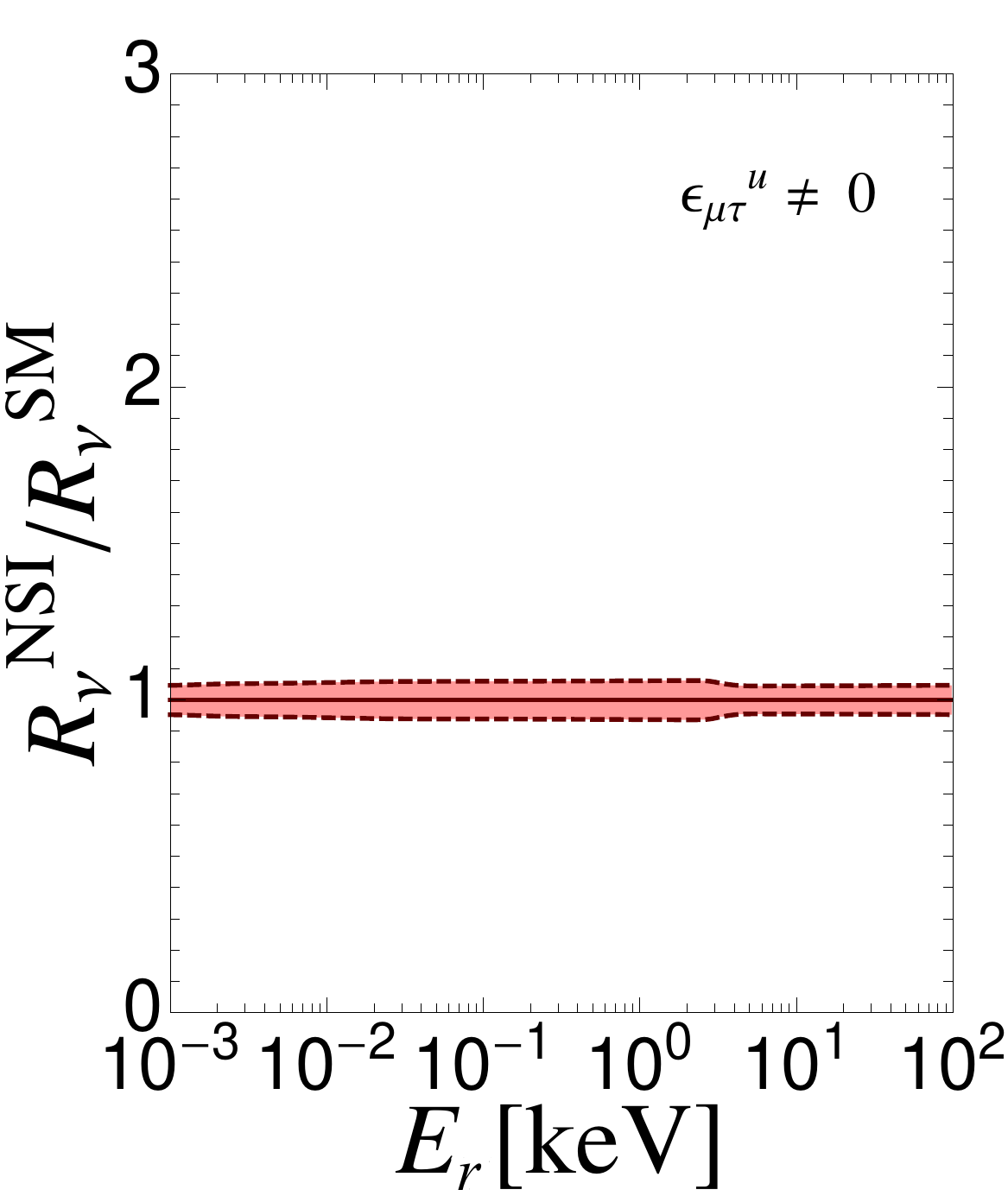}
  \includegraphics[scale=0.45]{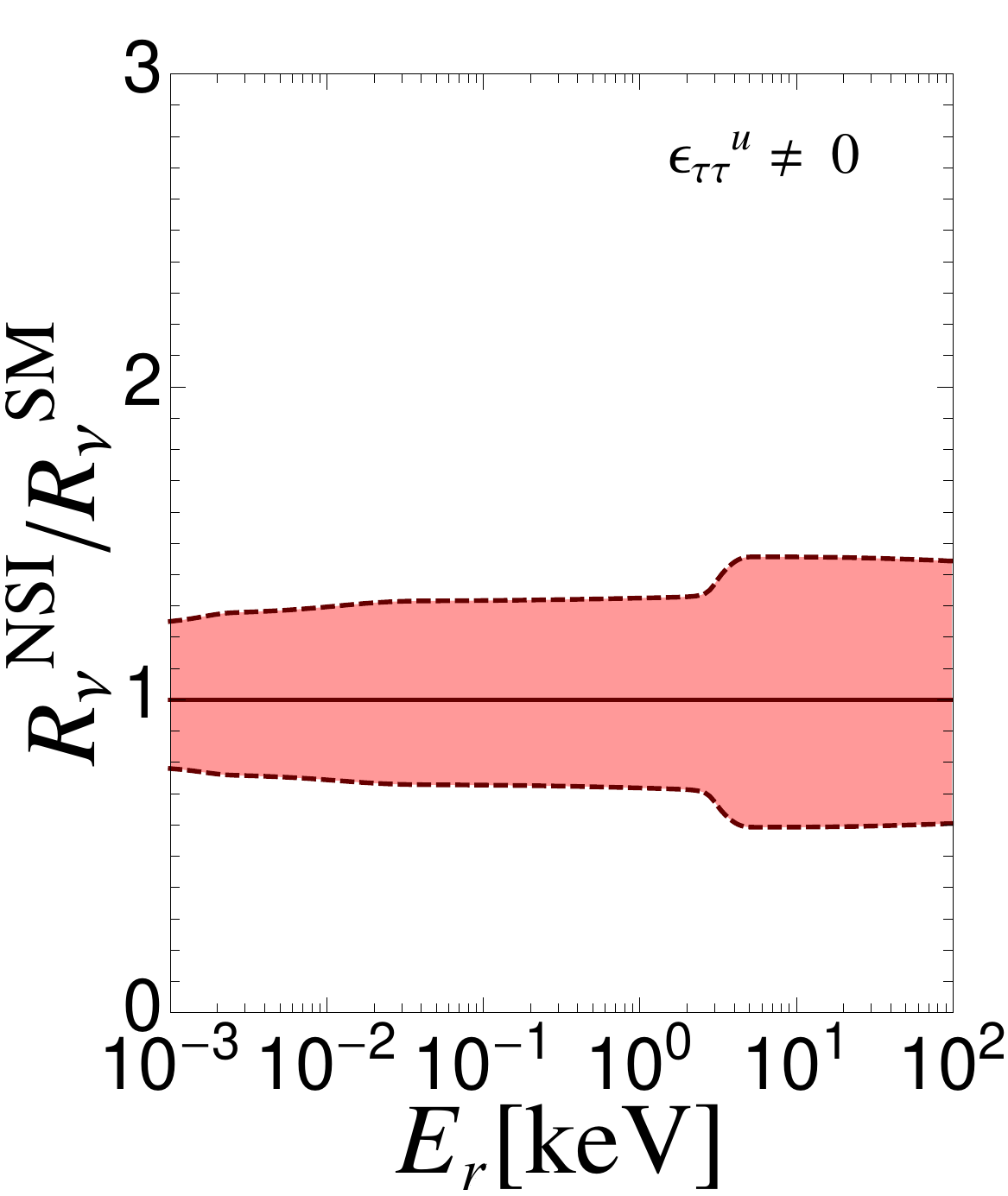}
  \caption{Expected number of neutrino-nucleus scattering events
    $R_\nu^\text{NSI}$ normalized to the SM expectation
    $R_\nu^\text{SM}$ as a function of recoil energy and for a
    Xe-based DM detector assuming an exposure
    $\mathcal{E}=1\,$~ton$\times$year and 100\% efficiency. The
    results include $pp$, $hep$, $^7$Be (ground and excited states),
    $^8$B, $pep$, $^{13}$N, $^{15}$O, $^{17}$F (CNO neutrinos) and
    atmospheric neutrinos fluxes (with $E_\nu< 100\,$~MeV).}
  \label{fig:number-of-events-NSI-DM}
\end{figure}

We now turn to the discussion of the impact of neutrino NSI on the
expected number of events in a Xe-based DM detector. For concreteness
we assume an exposure of
$\mathcal{E}=1\,\text{ton}\times\text{year}$. For practical reasons,
and using the same approach we used in the calculation of the averaged
survival probability in the Sun (see sec. \ref{sec:sol-atm-fluxes}),
we perform a single-parameter analysis, considering only up-type
couplings (results for down-type couplings resemble those found for
the up-quark case). We first calculate the neutrino-nucleus
differential event rate for ten equally-spaced values of the
corresponding NSI parameter:
$\epsilon_{ij}^u|_{a+1}=\epsilon_{ij}^u|_{a}+\delta \epsilon$, with
$\delta\epsilon=
\left(\epsilon_{ij}^u|_\text{max}-\epsilon_{ij}^u|_\text{min}\right)/10$.
From these results we then identify the parameters that
maximize/minimize the recoil spectra. And for these parameters we
calculate the neutrino-nucleus scattering events. For solar neutrinos
fluxes we employ the results derived in sec.~\ref{sec:sol-atm-fluxes},
eq.  (\ref{eq:event-rate-solar}) combined with the averaged survival
probabilities. For atmospheric neutrino fluxes we use the results from
sec.~\ref{sec:sol-atm-fluxes}. The results of our calculation are
displayed in fig. \ref{fig:number-of-events-NSI-DM}, which we regard
as the main result of this paper.

One can see that even those parameters that are forced to be small by
phenomenological constraints can have a sizable impact on the number
of events. Departures from the SM value (horizontal line at 1 in
fig. \ref{fig:number-of-events-NSI-DM}) are energy dependent and are
particularly pronounced in the solar-atmospheric transition, at
$E_r\simeq 5.8$~keV when $hep$ neutrinos reach their energy tail.
Note that in the absence of NSI propagation effects, the SM and NSI
neutrino-nucleus scattering rates differ only by a global numerical
factor, $Q_\text{SM}\to Q_{\nu_i}$ (see eq. (\ref{eq:QNSI-i})).  Thus,
this recoil energy dependence is a consequence of NSI propagation
effects.  Neutrino NSI not only lead to enhancements of the total
number of events but can as well produce depletions (as firstly noted
in ref. \cite{Dutta:2017nht}), with both being relevant for DM direct
detection experiments. The former places stronger limits on the
sensitivity one can reach in a multi-ton DM detector, while the latter
enables measurements of WIMP-nucleon cross sections that otherwise
would be challenging reaching.

Overall, one observes that no matter the coupling a deviation from the
SM expectation is always possible. However, diagonal couplings lead to
the largest deviations which can either be enhancements or depletions
of the event rate. Particularly important for solar neutrinos, are
deviations in the energy range where the background is dominated by
$^8$B, since it is in that range where solar neutrinos can mimic a
WIMP-nucleus signals \cite{Billard:2013qya}. In there, our findings
show that $\epsilon_{\mu\mu}^u$ leads to enhancements of
$R_\nu^\text{NSI}$ which can exceed the SM value by about 40\%. For
$\epsilon_{\tau\tau}^u$, enhancements are slightly smaller, but still
substantial. For $\epsilon_{ee}^u$ they reach values of order 2 or so,
being for this coupling the largest possible. This is somehow
expected, as is for this coupling that bounds are less severe. Note
that compared to what was found in ref. \cite{Dutta:2017nht} this
enhancement is larger, with the difference just due to the range
employed. In ref. \cite{Dutta:2017nht}
$\epsilon_{ee}^u|_\text{max}=0.45$, value derived from from
oscillation and deep inelastic scattering data \cite{Coloma:2017egw},
while in our case $\epsilon_{ee}^u|_\text{max}=0.6$. This difference,
although small, leads to a sufficiently large enhancement of the
neutrino-nucleus cross section (see
fig. \ref{fig:cross-sections-solar}) that accounts for the larger
deviation we found.

Since the largest deviations are found for diagonal NSI we focus in
these cases, discussing in a bit more detail the results for
$\epsilon_{ee}^u$ and $\epsilon_{\mu\mu}^u$. Maximization or
minimization of $R_\nu$ is determined by the way in which the NSI
parameters affect detection and propagation. The latter relevant only
in the solar sector. For $\epsilon_{ee}^u$ we found
$R_\nu^\text{NSI}|_\text{max}$ at $\epsilon_{ee}^u|_\text{max}=0.6$
for both, the region where recoils are controlled by solar neutrinos
fluxes ($E_r\lesssim 5.8$~keV) and the region governed by atmospheric
neutrinos ($E_r\gtrsim 5.8$~keV).  This can be understood as
follows. In the solar region $d\sigma_{\tilde \nu_\mu}/dE_r$ does not
depend on $\epsilon_{ee}^u$ and so it keeps always its SM value (see
fig. \ref{fig:cross-sections-solar}). The $\tilde \nu_e-N$ cross
section instead reaches its maximum exactly at
$\epsilon_{ee}^u|_\text{max}$. The survival probability behaves in the
opposite way, as $\epsilon_{ee}^u$ increases $\mathcal{P}_{ee}$
decreases globally, reaching minimum values exactly at
$\epsilon_{ee}^u|_\text{max}$ (see fig. \ref{fig:prob}, left
graph). This effect tends to deplete the contribution of the
$\tilde \nu_e-N$ scattering process to the total number of
events. However, since for $\epsilon_{ee}^u=0.6$ the survival
probability decreases at most $20\%$ from its SM value, the overall
effect is not sufficiently strong resulting in
$R_\nu^\text{NSI}|_\text{max}$ at $\epsilon_{ee}^u|_\text{max}$.
$R_\nu^\text{NSI}|_\text{min}$ is found at about the point where
$d\sigma_{\tilde \nu_e}/dE_r$ reaches its minimum,
$\epsilon_{ee}^u\simeq 0.2$. In the atmospheric sector, due
$\mathcal{P}_{e\mu}, \mathcal{P}_{e\tau}\ll 1$ the event rate is
mainly controlled by the $\nu_e-N$ cross section, which can be largely
modified by NSI effects (see figs. \ref{fig:atmospheric-prob} and
\ref{fig:atm-xsec}). This in turn is reflected in the non-depletion of
$R_\nu^\text{NSI}$ when passing from the solar to the atmospheric
regions.

The parameter dependence in the case of $\epsilon_{\mu\mu}^u$ is
rather different.  For solar neutrinos, $R_\nu^\text{NSI}|_\text{max}$
and $R_\nu^\text{NSI}|_\text{min}$ are found where the
$\tilde\nu_\mu-N$ cross section is maximized/minimized (see
fig. \ref{fig:cross-sections-solar}, right graph). When solar fluxes
are no longer relevant and atmospheric kick in, the maximum expected
number of events decreases. From (\ref{eq:atm-event-rates-elec}) one
can see that contributions from $\nu_e+\bar\nu_e$ fluxes are mainly
SM, sizable deviations are due only to $\nu_\mu+\bar\nu_\mu$
fluxes. They are however not as large as in the case of
$\epsilon_{ee}^u$ due to the sizable value of $\mathcal{P_{\mu\tau}}$
which tends to deplete the value of the second term in
eq. (\ref{eq:atm-event-rates-muon}), thus leading to small
$R_\nu^\text{NSI}|_\text{max}$.

The parameter dependence in the case $\epsilon_{\tau\tau}^u$ is rather
similar to the one found in the $\epsilon_{\mu\mu}^u$ scenario in the
solar sector, but in the atmospheric region the behavior is
different. From eqs. (\ref{eq:atm-event-rates-elec}) and
(\ref{eq:atm-event-rates-muon}) this is somehow
expected. Contributions from electron neutrino fluxes barely deviate
from SM values, substantial deviations can only arise from muon
neutrino fluxes, with the $\nu_\tau-N$ cross section weighted by the
$\nu_u-\nu_\tau$ oscillation probability which is of order $0.35$ (see
fig. \ref{fig:atm-xsec}). This results in slightly larger deviations
that those found in the $\epsilon_{\mu\mu}^u$ case. Non-diagonal NSI
couplings are subject to tighter constraints, thus resulting in
smaller effects. The exception being the case of the
$\epsilon_{e\tau}^u$ parameter, for which we found no depletion and
enhancements of order $40\%$, inline with what was reported in
\cite{Dutta:2017nht}.

The exact value (range) of the mediator mass for which our analysis is
valid depends---of course---on the recoil energy and hence on the
neutrino sector, solar or atmospheric. For the
$\epsilon_{\mu\mu}^u\neq 0$ case
(fig. \ref{fig:number-of-events-NSI-DM} left top panel), we show the
minimum mass that the mediator should have. In the solar sector masses
below $\sim 120$ MeV suffice, with the exact value depending on the
specific recoil energy. For recoil energies associated with
atmospheric neutrinos fluxes, $m_\text{med}\gtrsim 120\,$MeV. Thus,
while for solar neutrinos most of the mediator mass range is
available, for atmospheric neutrinos the analysis is limited to a
narrow window. However, as we have already pointed out is for this
window where NSI effects may be expected to be maximized.
\subsection{Implications for background-free sensitivities}
\label{sec:background-free-sens}
C$\nu$NES leads to recoil energy spectra which are closely degenerate
with that induced by WIMP-nucleus scattering. As a result, once a
direct detection experiment becomes sensitive to the neutrino
background there is a limit on the WIMP cross section
($\sigma_{\text{DM}-n}$) that such experiment can explore, due to the
inability to distinguish a neutrino- from a WIMP-induced signal. This
limit defines the so-called neutrino ``floor''.

With the C$\nu$NES process enhanced (diminished) by the
neutrino-nucleus NSI, one expects the neutrino floor to be more (less)
severe. Thus, in order to quantify the impact of NSI on the reach of
next-generation DM detectors, we have quantified the 90\% CL
sensitivities achievable. To do so, we have followed the method
introduced in ref. \cite{Billard:2013qya} which enables the
determination of the smallest WIMP cross section that can be measured
for any WIMP mass in the presence of one-neutrino background event. It
is worth pointing out that the limit on the WIMP cross section derived
this way is just an estimate of what one should really expect.  The
precise determination of the saturation point, where the WIMP cross
section one can access does not reduce with increasing (feasible)
exposure, requires abandoning the background-free paradigm and using
instead a likelihood analysis \cite{Billard:2013qya}. Here, however,
we aim just at a preliminary understanding of the possible impact that
NSI might have on direct detection sensitivities, and therefore stick
to a simple background-free analysis.

\begin{figure}
  \centering
  \includegraphics[scale=0.5]{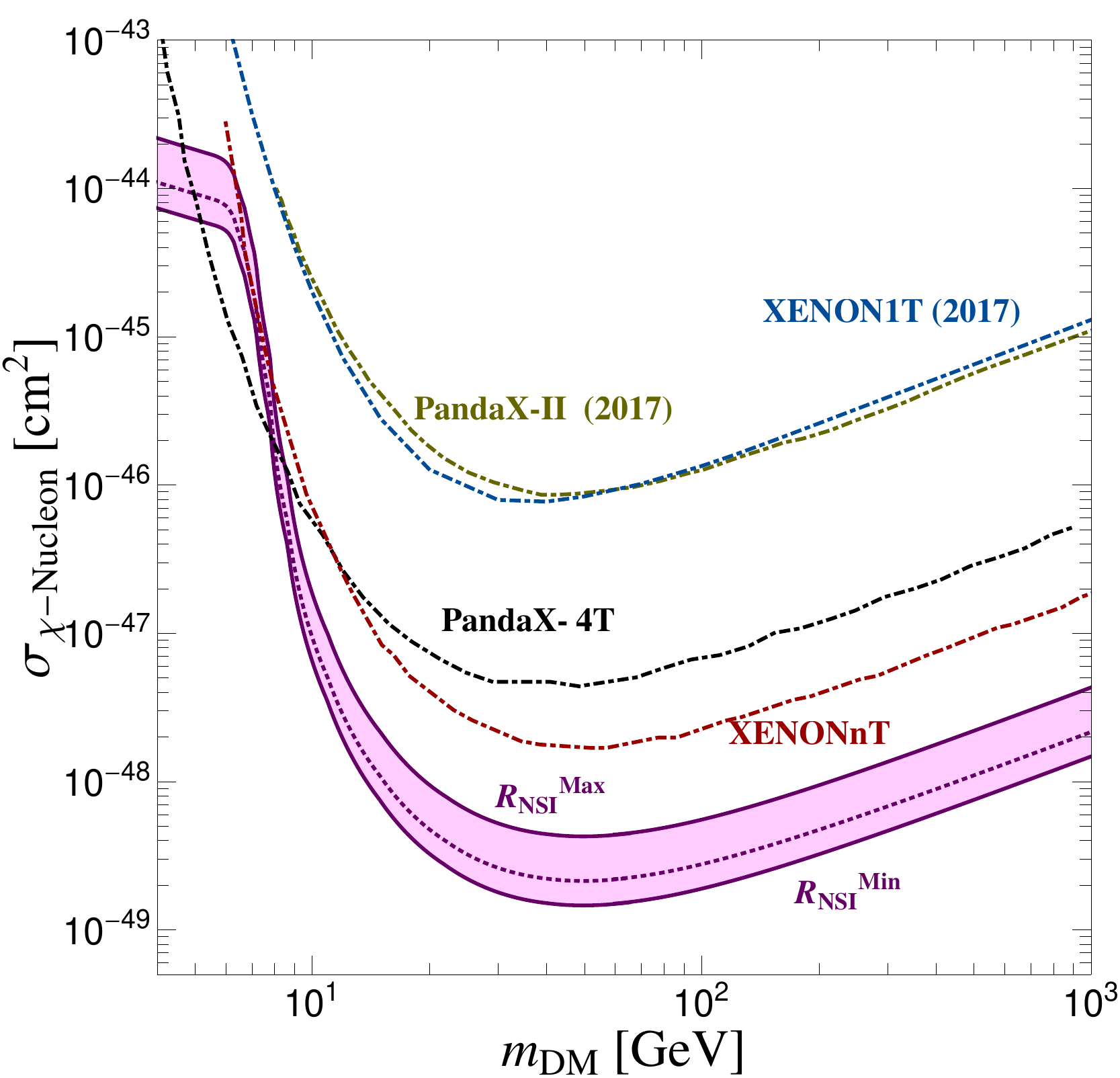}
  \caption{90\% CL background-free sensitivities in the presence of
    neutrino-nucleus NSI. The purple shaded region indicates the
    effects of NSI on the background-free sensitivities.  For
    comparison the SM 90\% CL background-free sensitivity curve is
    shown (dotted purple curve). These sensitivities have been derived
    by adjusting the exposure such that any measurement contains
    one-neutrino event \cite{Billard:2013qya}. It has been calculated
    for the case $\epsilon_{ee}^u\neq 0$.}
  \label{fig:background-free-sensitivities}
\end{figure}
We calculate the WIMP cross section as a function of the WIMP mass
by fixing the WIMP-nucleus scattering event rate, $R_\text{DM}$,
to 2.3 (90\% CL sensitivity limits), namely
\begin{equation}
  \label{eq:cross-section-as-a-fun-mDM}
  \sigma_{\text{DM}-n}=\frac{2.3}
  {\mathcal{E}(E_r^\text{th})
  \int_{E^\text{th}_r}^{E_r^\text{max}}
  \left . dR_\text{DM}/dE_r\right|_{\sigma_{\text{DM}-n}=1}}\ .
\end{equation}
Here the WIMP event rate, which depends upon astrophysical, nuclear
physics and particle physics parameters \cite{Lewin:1995rx} reads
\begin{equation}
  \label{eq:WIMP-event-rate}
  \frac{dR_\text{DM}}{dE_r}=\mathcal{E}\,
  \frac{\rho_\text{DM}\,\sigma_{\text{DM}-N}}{2\,m_\text{DM}\,\mu_N^2}
  F^2(E_r)
  \frac{k_0}{k}\frac{1}{(\pi\,v_0^2)^{3/2}}
  \,\int^{v_\text{esc}}_{v_\text{min}}\,
  e^{-|\boldsymbol{v}+\boldsymbol{v_\text{lab}}|^2/v_0^2}\,
  \frac{d^3\boldsymbol{v}}{|\boldsymbol{v}|}\ ,
\end{equation}
where $\mathcal{E}$ is the exposure, $\rho_\text{DM}$ is the DM
density in the local halo, $\sigma_{\text{DM}-N}$ is the DM-nucleus
cross section, $\mu_N$ is the DM-nucleus reduced mass, $v_0$ is the
most likely speed of the DM in the halo, $F^2(E_r)$ is the form factor
given in~(\ref{eq:helm-f-fac}) and $\boldsymbol{v}$ is the velocity of
the DM particles. The velocity distribution is Maxwellian and
truncated at the DM escape velocity:
$|\boldsymbol{v}+\boldsymbol{v_\text{lab}}|=\boldsymbol{v_\text{esc}}$,
with $\boldsymbol{v_\text{lab}}$ the laboratory (Earth) velocity
relative to the DM local halo. The factor $k_0/k$ is a normalization
factor obtained by integrating the Maxwellian DM velocity distribution
from 0 to $v_\text{esc}$ and is given by
\begin{equation}
  \label{eq:k0-over-k}
  \frac{k_0}{k}=
  \left[
    \text{erf}\left(\frac{v_\text{esc}}{v_0}\right)
    -
    \frac{2}{\sqrt{\pi}}\frac{v_\text{esc}}{v_0}e^{-v_\text{esc}^2/v_0^2}
  \right]^{-1}\ .
\end{equation}
Rather than integrating numerically the velocity distribution in
(\ref{eq:WIMP-event-rate}) we have used its exact analytic form,
namely \cite{Lewin:1995rx}
\begin{equation}
  \label{eq:velocity-distribution-integral}
  \int_{v_\text{min}}^{v_\text{esc}}\,
  e^{-|\boldsymbol{v}+\boldsymbol{v_\text{lab}}|^2/v_0^2}\,
  \frac{d^3\boldsymbol{v}}{|\boldsymbol{v}|}
  =
  2\pi\,v_0^2\,
  \left\{
    \frac{\sqrt{\pi}}{4}\frac{v_0}{v_\text{lab}}
    \left[
    \text{erf}\left(\frac{v_\text{min}+v_\text{lab}}{v_0}\right)
    -
    \text{erf}\left(\frac{v_\text{min}-v_\text{lab}}{v_0}\right)
    \right]
    -
    e^{-v_\text{esc}^2/v_0^2}
  \right\}\ ,
\end{equation}
with $v_\text{min}=\sqrt{2E_r/m_\text{DM}\,r}\,\times c$ and
$r=4m_\text{DM}m_N/(m_\text{DM}+m_N)^2$. We have written as well the
DM-nucleus cross section in terms of the DM-nucleon cross section
\begin{equation}
  \label{eq:dm-nucleus-dm-nucleon-x-secs}
  \sigma_{\text{DM}-N}=\frac{\mu_N^2}{\mu_n^2}\,A^2\,
  \sigma_{\text{DM}-n}\ ,
\end{equation}
with $\mu_n$ the DM-nucleon reduced mass. For the different
astrophysical quantities we have used: $\rho_\text{DM}=0.3$
GeV/c$^2$/cm$^3$, $v_0=220$ km/s, $v_\text{lab}=232$ km/s and
$v_\text{esc}=544$ km/s. For the integration we varied $E_r^\text{th}$
from $10^{-3}$ to $10^2$ keV (with $10^3$ partitions) and took
\begin{equation}
  \label{eq:maximum-recoil-energy}
  E_r^\text{max}=\frac{m_\text{DM}}{2}
  \left(\frac{v_\text{esc}+v_\text{lab}}{c}\right)^2
  \frac{r}{2}\ ,
\end{equation}
with $r$ given after
eq. (\ref{eq:velocity-distribution-integral}). The exposure has been
tuned so to assure that each $E_r^\text{th}$ isocontour in the
$\sigma_{\text{DM}-n}-m_\text{DM}$ plane contains one-neutrino
event. This has been done by calculating the exposure as follows:
\begin{equation}
  \label{eq:Exposure}
  \mathcal{E}(E^\text{th}_r)=\frac{1\,\text{neutrino event}}
  {\int_{E^\text{th}_R}dR_\nu/dE_r}\ .
\end{equation}

The result is shown in fig. \ref{fig:background-free-sensitivities},
where we have plotted the SM sensitivities as well and included
current (future) exclusion limits from XENON1T and PandaX-II
\cite{Aprile:2017iyp,Cui:2017nnn} (XENONnT and PandaX-4T
\cite{pandax4T}). The sensitivities shown in the plot refer to those
obtained by calculating the exposure in (\ref{eq:Exposure}) for the
values of $\epsilon_{ee}^u$ that maximize/minimize $dR_\nu/dE_r$
(lower and upper curves in fig. \ref{fig:number-of-events-NSI-DM},
left-top graph). The result in
fig. \ref{fig:background-free-sensitivities} show that the effects of
neutrino-nucleus NSI, if present, work either way. They can worsen
sensitivities or improve them. Note, that despite the effect being
small it might have consequences. For example, if the NSI-induced
depletion is effective (lower purple curve) the background-free
sensitivities at XENONnT will be less affected. If instead the
NSI-induced enhancement is present, XENONnT or PandaX-4T could start
measuring neutrino scattering events sooner than expected.
\section{Conclusions}
\label{sec:concl}
The identification of a DM signal in next-to-next generation
(multi-ton scale) detectors will be seriously challenged by
irreducible solar and atmospheric neutrino backgrounds. Identifying
all the features of this background is therefore mandatory to
determine the implications it will have and the challenges it will
pose. With this in mind, in this paper---following
\cite{Dutta:2017nht}---we have entertained the possibility that
neutrinos are subject to NSI and have studied the impact they would
have on the expected number of neutrino-nucleus scattering events in a
Xe-based detector assuming an exposure of
$1\;\text{ton}\cdot\text{year}$. We have as well determine the
background-free sensitivities achievable if these interactions are
present, showing that they can either worsen or improve them, although
only slightly in both cases.

Despite the recent observation of the C$\nu$NES process by the
COHERENT experiment and its consistency with the SM expectation, NSI
couplings can still be relatively large, even when the COHERENT
scattering data is combined with data from neutrino oscillation
experiments. Thus, by using current constraints applicable in
scenarios where neutrino NSI arise from mediators with masses in the
range $[10, 10^3]\,\text{MeV}$, we have calculated the expected number
of neutrino-nucleus scattering events in the presence of NSI including
propagation and detection effects and extending the analysis over the
solar and atmospheric neutrino sectors.  In the former case, we have
included all $pp$ and CNO solar neutrino fluxes, while in the latter
case we have included $\nu_e, \nu_\mu, \bar\nu_e$ and $\bar\nu_\mu$
fluxes for neutrino energies up to $\sim 100$ MeV. To simplify the
computational burden, we have carried out a single-parameter analysis
in which all couplings were put to zero but one and repeated the
analysis for the remaining NSI couplings. Since we used constraints
from \cite{Coloma:2017ncl}, we analyzed up- and down-quark couplings
independently and presented only the results for the former, as the
results for the latter are rather similar.

For the solar sector, we have studied propagation in the mass
dominance limit and have taken into account the distribution of
neutrino production within the Sun, for the calculation of the
averaged neutrino survival probability. In the atmospheric region,
Earth matter effects are negligible due to the largest relevant
neutrino energies being below $\sim 100$ MeV. In this sector then, NSI
affect the expected number of neutrino-nucleus scattering events only
through detection. We have anyway included vacuum oscillation
probabilities (properly integrated over the zenith angle) in the
determination of the number of scattering events.

Our main results are given by fig. \ref{fig:number-of-events-NSI-DM}
and fig. \ref{fig:background-free-sensitivities} and can be summarized
as follows. Largest deviations from the SM expectation are more
pronounced for flavor-diagonal NSI and can operate either enhancing
the number of events or depleting them. For the most relevant recoil
energy interval in the solar sector (that in which the neutrino flux
is determined by $^8$B neutrinos in which the background can mimic
WIMP-nucleon scattering events) the largest enhancement is found for
$\epsilon_{ee}^u$ and can exceed the SM expectation up to a factor
$\sim 2$. For $\epsilon_{\mu\mu}^u$ and $\epsilon_{\tau\tau}^u$,
enhancements are smaller but still sizable. Depletions in that energy
interval can amount to about 30-40\% for all diagonal NSI
parameters. Moving to higher neutrino energies we found that when
atmospheric neutrino fluxes dominate the background, enhancements as
large as twice the SM expectation are possible when $\epsilon_{ee}^u$
is present, in all other cases they are below this value but still are
sizable. Depletions in the atmospheric sector can amount to about
$50$\% if $\epsilon_{\tau\tau}^u$ is at work. For the off-diagonal
couplings $\epsilon_{e\mu}^u$ and $\epsilon_{e\tau}^u$ we did not find
any depletion, but we did for $\epsilon_{\mu\tau}^u$ although rather
small (order $5\%$) and so probably not experimentally
measurable. Thus, if the neutrino-nucleus event rate for energies
dominated by atmospheric neutrino fluxes is found to be below SM
expectations---arguably---that will support the presence of new
physics featuring flavor-diagonal NSI, assuming one can disentangle the
new interactions above any other effect.

Deviations, though small, might have an effect on the sensitivities
achievable. To estimate the extent of such effects we have calculated
the background-free sensitivities. We have shown that neutrino-nucleus
NSI, if present, may worsen sensitivities or improve them. The effect
is small, but might have observable consequences. We have stressed
that if neutrino NSI are present either XENONnT or PandaX-4T, among
others, will start observing neutrino-induced nuclear recoil events
sooner than expected.
\section*{Acknowledgments}
D.A.S. would like to thank Atri Bhattacharya, Jean-Rene Cudell, Thomas
Hambye, Julian Heeck, Concha Gonzalez-Garcia, Florica Stancu and
Mariam Tortola. In particular, Concha Gonzalez-Garcia for discussions
regarding refs. \cite{Coloma:2017egw,Gonzalez-Garcia:2013usa}, Mariam
Tortola for discussions on neutrino flavor oscillations in the Sun and
Jean-Rene Cudell and Florica Stancu for useful conversations on
nuclear physics. He would like also to thank the ``Service de Physique
Th\'eorique'' of the ``Universit\'e Libre de Bruxelles'' and the STAR
research unit of the ``Universit\'e de Li\`ege'' for the warm
hospitality during the early stages of this work and partial financial
support. This research has been supported by F.R.S.-FNRS through the
program ``Bourses de S\' ejour Scientifique'' and by the Chilean grant
``Unraveling new physics in the high-intensity and high-energy
frontiers'', Fondecyt No 1171136. NR was funded by proyecto FONDECYT
Postdoctorado Nacional (2017) num. 3170135.  The work of M.T.  is
supported by the FNRS-FRS, the Belgian Federal Science Policy Office
through the Interuniversity Attraction Pole P7/37 and the Belgian
Institute Interuniversitaire des Sciences Nucl\'eaires (IISN).
\bibliography{references}
\end{document}